\documentclass[superscriptaddress,reprint, aps, pra]{revtex4-2}

\usepackage{subfigure}

\usepackage{amsmath}

\usepackage{enumitem}

\usepackage{graphicx}

\graphicspath{{images/}}

\usepackage{xcolor}
\usepackage{soul}
\usepackage[normalem]{ulem}

\usepackage{dcolumn}

\usepackage{bm}

\usepackage[colorlinks=true,allcolors=blue]{hyperref}

\begin{document}

\title{Non-orthogonal qubit states expansion for the asymmetric quantum Rabi model }

\author{Zi-Min Li}
\affiliation{Department of Theoretical Physics, Research School of Physics, Australian National University, Canberra ACT, 2601, Australia}

\author{Devid Ferri}
\affiliation{Department of Theoretical Physics, Research School of Physics, Australian National University, Canberra ACT, 2601, Australia}

\author{Murray T. Batchelor}
\email{batchelor@cqu.edu.cn}
\affiliation{Centre for Modern Physics, Chongqing University, Chongqing 400044, The People's Republic of China}
\affiliation{Department of Theoretical Physics, Research School of Physics, Australian National University, Canberra ACT, 2601, Australia}
\affiliation{Mathematical Sciences Institute, Australian National University, Canberra ACT, 2601, Australia}

\date{\today}


\begin{abstract}
We present a physically motivated variational wave function for the ground state of the asymmetric quantum Rabi model (AQRM). 
The wave function is a weighted superposition of squeezed coherent states entangled with non-orthogonal qubit states, 
and relies only on three variational parameters in the regimes of interest where the squeezing effect becomes negligible.
The variational expansion describes the ground state remarkably well in almost all parameter regimes, especially with arbitrary bias. 
We use the variational result to calculate various relevant physical observables of the ground state, and make a comparison with 
existing approximations and the exact solution. 
The results show that the variational expansion is a significant improvement over the existing approximations for the AQRM. 
\end{abstract}

\maketitle

\section{Introduction}

The quantum Rabi model (QRM), consisting of a two-level system interacting with a single mode bosonic field, 
describes the simplest interaction between light and matter \cite{Rabi_1936,*Rabi_1937}. 
The QRM has found applications in various fields, such as quantum optics \cite{Scully_1997,Fox2006} and condensed-matter physics \cite{Wagner1986}. 
For many decades, the QRM has continued to inspire developments from both the physical and mathematical perspectives \cite{Braak_2016}.
In the past decade, the rapid development of experimental techniques has given rise to the realization of the QRM in different 
experimental platforms, 
most notably in circuit quantum electrodynamics (cQED) \cite{Wendin_2017,Forn_D_az_2019,Frisk_Kockum_2019,Blais2020}. 
These developments place the QRM, along with its various generalizations, in the midst of the emerging technology of quantum devices.

Despite the simple form and long history of the QRM, the model was not solved until recently \cite{Braak_2011,Chen2012,Zhong_2013,Zhong_2014,Maciejewski2014,*Maciejewski_2014,Xie_2017,Braak_2019}. 
The exact solutions obtained from various methods all rely on the zeros of transcendental functions, 
making it hard to extract exact information, and to some extent, physical intuition from them. 
For this reason approximations to the QRM are still valued and thus actively sought. 
The most well known approximation in the long history of the QRM is the rotating-wave approximation (RWA), 
which leads to the Jaynes-Cummings model (JCM) \cite{Jaynes_1963}. 
The RWA omits the energy non-conserving terms, which are the counter-rotating terms, from the interaction Hamiltonian, 
and thus reduces the system Hamiltonian to infinitely many $2\times 2$ block matrices, which is then easily solvable. 
The JCM successfully describes the light-matter system in the strong coupling regime \cite{Fox2006}, 
where the coupling strength is much smaller than the cavity and atomic frequencies, but is much larger than the dephasing and decoherence rates. 
However, the RWA no longer holds in the ultra-strong coupling (USC) \cite{Forn_D_az_2019,Frisk_Kockum_2019} and deep-strong coupling (DSC) \cite{Casanova_2010,Rossatto_2017} regimes, which have recently been achieved in experiments \cite{Forn_D_az_2010,Yoshihara_2016,Yoshihara_2018}.

Various approximation schemes have been proposed to describe the QRM in the USC and DSC regimes  \cite{Feranchuk_1996,Irish_2005,Irish_2007,Gan_2010,Boite2020}.
Among them is the generalized rotating-wave approximation (GRWA) proposed by Irish \cite{Irish_2007}, 
which provides a framework to treat the QRM and related models. 
The GRWA is carried out by performing a displacement transformation \cite{Irish_2005} 
before exploiting the RWA, hence being called the {`generalized'} RWA. 
The GRWA works surprisingly well with arbitrary large coupling strength in near resonant cases. 
It has been further generalized by introducing a variational displacement parameter \cite{Zhang_2011,Yu_2012}, 
called the generalized variational method (GVM), and by introducing squeezing effects \cite{Zhang_2016,Zhang_2017}, 
called the generalized squeezing rotating wave approximation (GSRWA). 
The key point of all these approximations is to unitarily transform the QRM and discard terms corresponding to the counter-rotating terms, 
such that the final form is similar to the JCM, which can then be solved analytically.

The GRWA, as well as its further generalizations, have been applied to other light-matter interaction models.
The example of interest here, and most relevant to the current cQED experiments, 
is the asymmetric quantum Rabi model (AQRM) \cite{Zhang_2013,Mao_2018,Xie2020}.
The AQRM is a generalization of the QRM \cite{Li_2015,*Li_2016,Wakayama_2017,Kimoto_2020,Li2020a,Li2020} with a bias term, 
which naturally appears in the cQED setup to describe qubits with an asymmetric potential such as flux qubits \cite{Forn_D_az_2010,Yoshihara_2016,Yoshihara_2018,Forn_D_az_2019,Blais2020}.
The approximating treatments to the AQRM may be classified into the GRWA \cite{Zhang_2013}, 
the GVM \cite{Mao_2018} and the GSRWA \cite{Xie2020}, 
following the aforementioned methods. 
The work by Zhang {\it et al.} \cite{Zhang_2013} is a straightforward application of the GRWA, 
while Refs.~\cite{Mao_2018,Xie2020} further introduce the variational displacement and squeezing parameters, respectively. 
These three approximations are seen to describe the ground state of the AQRM reasonably well in specific parameter regimes. 
However, we notice that none of them performs satisfactorily in the relatively large bias cases, 
where the bias is comparable to the field frequency $\omega$ or qubit transition rate $\Delta$. 
This parameter regime is frequently used in cQED experiments.

In this paper, we introduce a variational wave function to describe the ground state of the AQRM.
We start from physical observations of the system evolution with respect to the coupling strength, 
and propose a trial wave function consisting of a weighted superposition of squeezed coherent states 
entangled with non-orthogonal qubit states \cite{Irish_2014,Leroux_2017}. 
The resulting trial function, depending on only three variational parameters, 
describes the system ground state remarkably well in almost all parameter regimes. 
A fourth variational parameter, based on the squeezing parameter, can also be introduced, but does not make any noticeable difference 
to the eigenvalues.
In particular, our Ansatz exhibits excellent agreement with the exact numerical solutions for arbitrary bias, which is crucial to the AQRM.

This paper is organized as follows. 
In section \ref{SectionNOQ} we make several physical observations, 
based on which we propose the Ansatz to describe the ground state of the AQRM. 
We calculate the ground state eigenvalues and some physical observables with this Ansatz and compare them with the exact numerical solutions, 
as well as with the known approximations, in section \ref{SectionProperties}. 
Section \ref{SectionDiscussion} is devoted to further discussion, including the validity of the variational Ansatz and 
its correspondence to other methods. 
Finally, a brief summary, with an outlook to future directions, is given in section \ref{SectionConclusion}.

\section{Non-orthogonal qubit states ansatz}\label{SectionNOQ}

\subsection{Model Hamiltonian}

The Hamiltonian of the AQRM reads ($\hbar=1$) 
\begin{equation}\label{HAQRM}
H = \dfrac{\Delta}{2}\sigma_z + \omega a^\dagger a + g \left(a^\dagger + a\right)\sigma_x + \dfrac{\epsilon}{2}\sigma_x, 
\end{equation}
where $\sigma_x$ and $\sigma_z$ are Pauli matrices for a two-level system with level splitting $\Delta$. 
The single mode bosonic field is described by the creation and annihilation operators $a^\dagger$ and $a$, and frequency $\omega$. 
The interaction between the two systems is via the coupling $g$. 
Regarding the Hamiltonian of the AQRM, there is another form often used, 
which effectively resorts to exchanging the Pauli matrices $\sigma_x$ and $\sigma_z$. 
Formally we can think of this unitary transformation to take form as a Hadamard transform. 
Alternatively, by redefining $\Delta$ as $-\Delta$, 
the other conventional form of the Hamiltonian can be obtained through a similar transform 
but with a unitary operator of the form $U = e^{\mathrm{i} \frac{\pi}{4} \sigma_y}$.
Throughout this paper we shall work with the Hamiltonian defined in Eq.~(\ref{HAQRM}). 

To gain some intuition of the ground state of the AQRM, 
we first consider the symmetric case $\epsilon=0$, i.e., the standard QRM, with 
\begin{equation}\label{HQRM}
H_R = \dfrac{\Delta}{2}\sigma_z + \omega a^\dagger a + g \left(a^\dagger + a\right)\sigma_x. 
\end{equation}
The QRM possesses parity symmetry \cite{Braak_2019}, namely $[P,H_R]=0$ with $P=\sigma_z(-1)^{a^\dagger a}$.

The QRM has two known limits in which the system can be easily solved.

\subsection{Zero coupling limit $g=0$}
The first case is $g=0$, where the system is decoupled to the field part and the qubit part, with the effective Hamiltonian now being
\begin{equation}\label{Hg0}
H_R^{g=0} =\omega a^\dagger a + \frac{\Delta}{2}\sigma_z,
\end{equation}
which can be diagonalized by the trivial tensor products $|n\rangle\otimes|\pm z\rangle$. 
Here $|n\rangle$ denotes the standard Fock states and $|\pm z\rangle$ are the eigenstates of $\sigma_z$. 
In this limit, the ground state simply reads
\begin{equation}\label{GSg0}
\psi_0^{g=0} = |0\rangle\otimes|\!-z\rangle. 
\end{equation}

\subsection{Large coupling limit $g/\Delta\rightarrow\infty$}
In the limit $g/\Delta\rightarrow\infty$, the system Hamiltonian is
\begin{equation}\label{Hlargeg}
H_R^{\Delta=0} = \omega a^\dagger a + g\sigma_x \left(a^\dagger + a\right),
\end{equation}
which can be regarded as two displaced harmonic oscillators \cite{Irish_2005}. 
This limit is achieved either with $g/\omega\rightarrow\infty$, but $\Delta/\omega$ being finite, or simply $\Delta=0$. 
For the $\Delta=0$ case, the physical interpretation of the system is two harmonic oscillators displaced in different directions, 
sharing no tunnelling between each other. 
On the other hand, if $g/\omega\rightarrow\infty$ and $\Delta/\omega$ is finite, physically this means that the two oscillators are so far 
away that finite $\Delta/\omega$ cannot induce any tunnelling.  
Therefore, these two limit cases are physically equivalent. 

The Hamiltonian (\ref{Hlargeg}) is diagonalized by a bosonic displaced transformation \cite{Irish_2005}
\begin{equation}\label{DisplacementTransformation}
\mathcal{D}(\alpha) = e^{\alpha(a^\dagger-a)},
\end{equation}
with displacement amplitude $\alpha=g/\omega$. 
The eigenstates of the displaced oscillators are obtained as
\begin{equation}\label{key}
\psi_{n}^\text{do} = | \! \pm\alpha, n\rangle\otimes | \! \mp x \rangle, 
\end{equation}
where $| \! \pm\alpha, n\rangle = \mathcal{D}(\pm\alpha)|n\rangle$ are the displaced Fock states. 
The displacement directions are determined by the qubit states $| \! \mp x \rangle$. 
Note that the states $|\! \pm g/\omega, 0\rangle\otimes | \!  \mp x \rangle$ are doubly degenerate, 
and the ground state with definite parity can be expressed as
\begin{equation}\label{Largeg}
\psi_0^{\Delta=0} = \dfrac{1}{\sqrt{2}}\left(|\alpha\rangle\otimes|\!-x\rangle - |\!-\alpha\rangle\otimes|\!+x\rangle\right),
\end{equation}
in which $|\alpha\rangle$ denotes the displaced vacuum state.
The negative parity of Eq.~(\ref{Largeg}) is chosen to be consistent with the ground state of the QRM in general cases.

\subsection{Intermediate coupling regime}
Having discussed the easily solvable limits, we now consider the intermediate regime. 
It is natural to think about the evolution of the system with respect to the coupling strength $g$, 
given that we have exact solutions in the two limits. 
If we increase the coupling strength $g$ from 0 to $\infty$, the qubit of the ground state evolves from $|\!-z\rangle$ to $|\!\pm x \rangle$. 

\begin{figure}[htpb]\centering      
	\subfigure[]{                
		\centering                                                 
		\includegraphics[width=.7\linewidth]{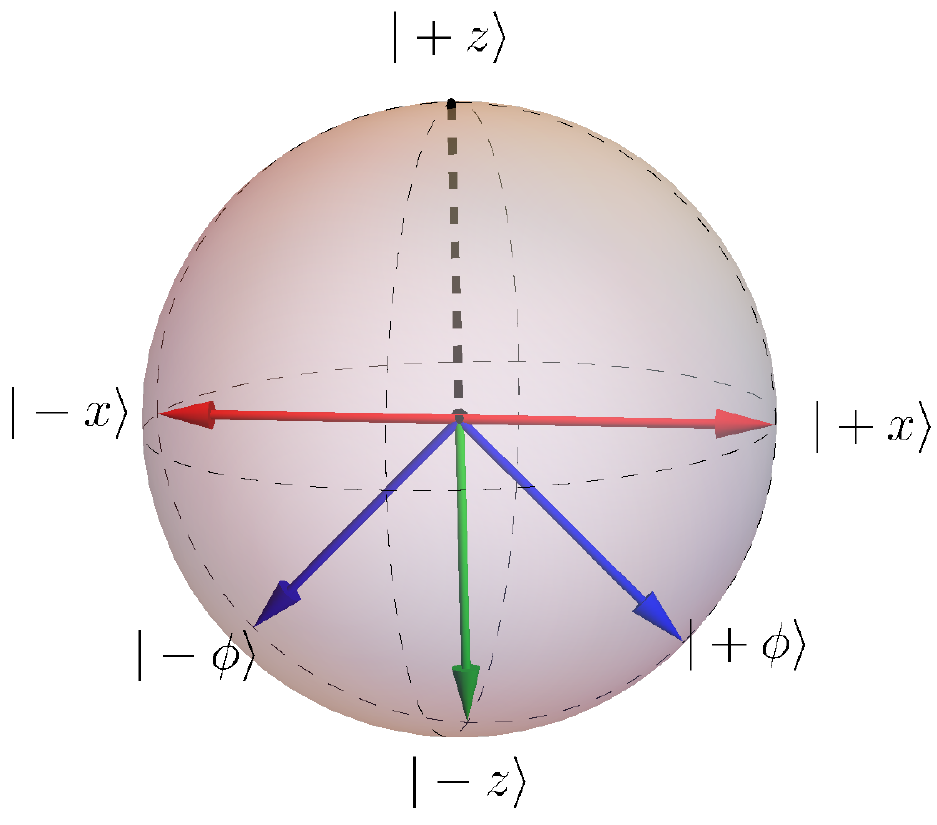}        
		\label{BlochSphere}
	}         
	\subfigure[]{                
		\centering                                                 
		\includegraphics[width=.46\linewidth]{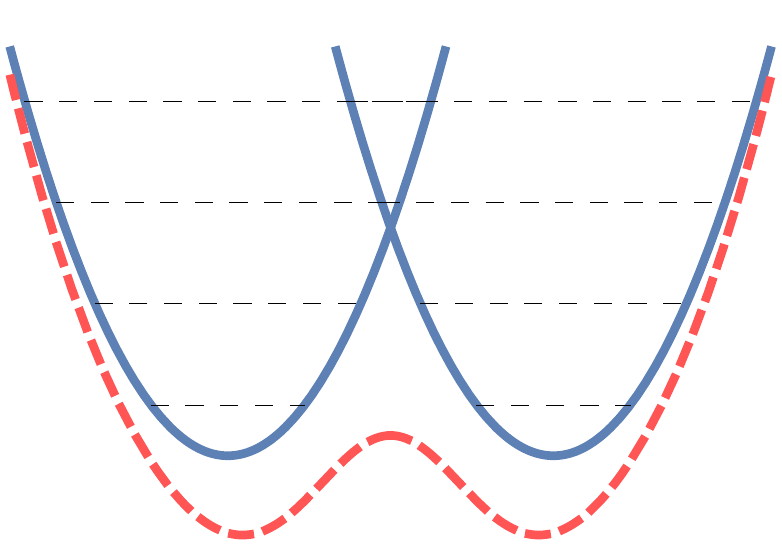}        
		\label{EffectivePotentialsQRM}
	}   
	\subfigure[]{                
		\centering                                               
		\includegraphics[width=.46\linewidth]{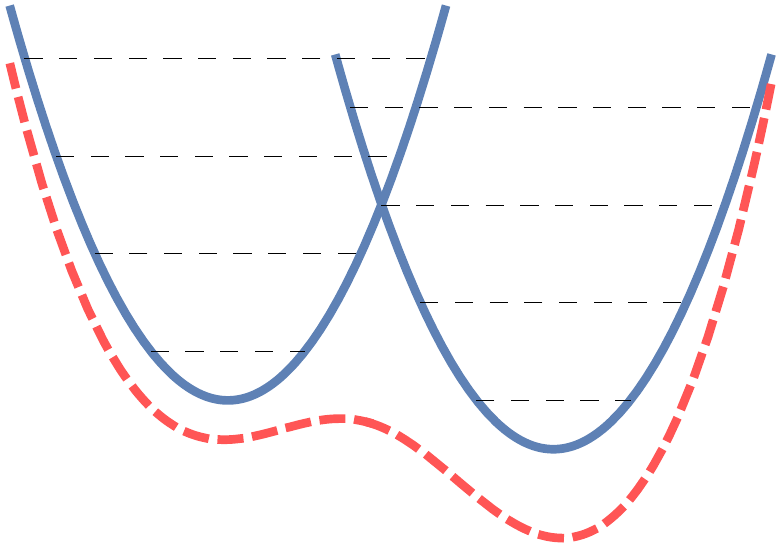}        
		\label{EffectivePotentialsAQRM} 
	}                
	\caption{(a) Schematic qubit states on the Bloch sphere. Regarding the ground state, when $g=0$, the qubit is in the state $|\!-z\rangle$, and in the limit $g\rightarrow\infty$, the qubit is in a superposition of the states $|\!\pm x\rangle$. In the intermediate coupling regimes, the qubit takes the states $ |\!\pm\phi\rangle $ in between.  (b) Schematic effective potentials of the QRM. In the intermediate coupling regimes, non-zero $\Delta$ induces tunnelling between the two displaced oscillators, leading to anharmonicity and smaller displacements. The resulting effective potential is symmetric with respect to the $y$-axis, indicating the existence of parity symmetry. (c) Schematic effective potentials of the AQRM. Non-zero bias $\epsilon$ introduces energy shifts to the oscillators, leading to asymmetry of the combined effective potential.} 
	\label{EffectivePotentials}
\end{figure}

A reasonable description for the qubit ground state in the intermediate regime would be of some spin states in between, 
as depicted in Fig.~\ref{BlochSphere}. 
The qubit state may be regarded as a superposition of two spin states obtained through rotating 
$|\!+z\rangle$ by an angle $\pm\theta$ with respect to the $y$-axis. 
When $g=0$, this angle takes the value $\theta=\pi$, and the qubit sits in the state $|\!-z\rangle$. 
In the limit $g\rightarrow\infty$, the state is rotated by $\pm\theta = \pm \frac{\pi}{2}$, and the qubit is in a superposition of $|\!\pm x \rangle$. 

On the other hand, tunnelling between the two displaced oscillators is induced by non-zero $\Delta$ in the intermediate regime, 
such that the effective potentials tend to get closer \cite{Irish_2014,Ying_2015,Xie2020}, see Fig.~\ref{EffectivePotentialsQRM}. 
As a result, the displacement amplitude is no longer fixed as $g/\omega$, but should be a free parameter. 
Meanwhile, due to the tunnelling effect, the two displaced oscillators are no longer harmonic, but come with some non-trivial anharmonicity. 
Effectively, the field frequencies have changed and can be described by a squeezing parameter.

We now consider the full AQRM and switch on the bias term $\frac12 {\epsilon}\,\sigma_x$, 
which breaks the parity symmetry in the standard QRM. 
From the perspective of effective potentials, this asymmetric term leads to energy shifts to the oscillators, 
as displayed in Fig.~\ref{EffectivePotentialsAQRM}. 
The contributions to the ground state wave function from the two oscillators are therefore now unequal and 
generally the wave functions of the AQRM do not have any parity \cite{Ashhab_2020}. 

\subsection{The non-orthogonal qubits Ansatz}

With the above observations and analysis, it is now straightforward to write down an 
unnormalized trial function for the ground state of the AQRM, namely
\begin{equation}\label{NOQ}
\psi_0 = p \, |\alpha,\gamma\rangle\otimes|\phi_-\rangle - \sqrt{1-p^2} \, | \!-\alpha,\gamma\rangle\otimes|\phi_+ \rangle,
\end{equation}
where the weight $p$ generally breaks the parity of the wave function. 
The field is described by standard squeezed coherent states \cite{Scully_1997}
\begin{equation}\label{SCS}
|\!\pm\alpha,\gamma\rangle = \mathcal{S}(\gamma)\mathcal{D}(\pm\alpha)|0\rangle. 
\end{equation}
Here the displacement operator has been given in Eq.~(\ref{DisplacementTransformation}), and the squeezing operator is \cite{Scully_1997,Fox2006}
\begin{equation}\label{SqueezingOperator}
\mathcal{S}(\gamma) = e^{-\frac{\gamma}{2}\left({a^\dagger}^2 - a^2\right)}, 
\end{equation}
with unfixed squeezing parameter $\gamma$. 
The qubit is described by rotated spin states
\begin{equation}\label{NOQSpin}
|\phi_\pm \rangle = \cos\dfrac{\theta}{2} \, | \! +z \rangle \pm \sin\dfrac{\theta}{2} \, | \!-z \rangle ,
\end{equation}
which generally are not orthogonal to each other \cite{Irish_2014,Leroux_2017}. 
We thus refer to Eq.~(\ref{NOQ}) as the non-orthogonal qubits (NOQ) Ansatz.

There are four free parameters: $\alpha$, $\theta$, $\gamma$ and $p$, in the NOQ Ansatz (\ref{NOQ}).
For practical reasons, we would like to keep the number of variational parameters to be as few as possible, without sacrificing accuracy. 
In the limit $\Delta/\omega\rightarrow \infty$, a superradiance phase transition occurs in the QRM \cite{Hwang_2015,Chen2020}, 
accompanied by the spontaneous breaking of parity symmetry, and the squeezing effect is dominant \cite{Hwang_2010}. 
However, in the AQRM with nonzero $\epsilon$, the parity symmetry is already broken, and no quantum phase transition is known to take place.
Additionally, cQED experiments generally operate with negative detuning ($\Delta/\omega<1$) \cite{Fedorov_2010,Forn_D_az_2010,Yoshihara_2018}. 
We thus focus on the parameter regimes where $\Delta/\omega$ is not too large. 
Numerical observations show that in the regimes of interest, the squeezing effect becomes negligible. 
We therefore safely set $\gamma=0$ in the following calculations, and simply write the field states as 
$|\alpha_\pm\rangle = | \! \pm \alpha , \gamma \rangle$. 
As a result, we are left with only the three variational parameters $(\alpha,\theta,p)$.

We now turn to tests of the NOQ Ansatz Eq.~(\ref{NOQ}).

\section{Physical properties in the ground state}\label{SectionProperties}

\begin{figure}[t]\centering            
	\subfigure{                
		\centering                                                 
		\includegraphics[width=\linewidth]{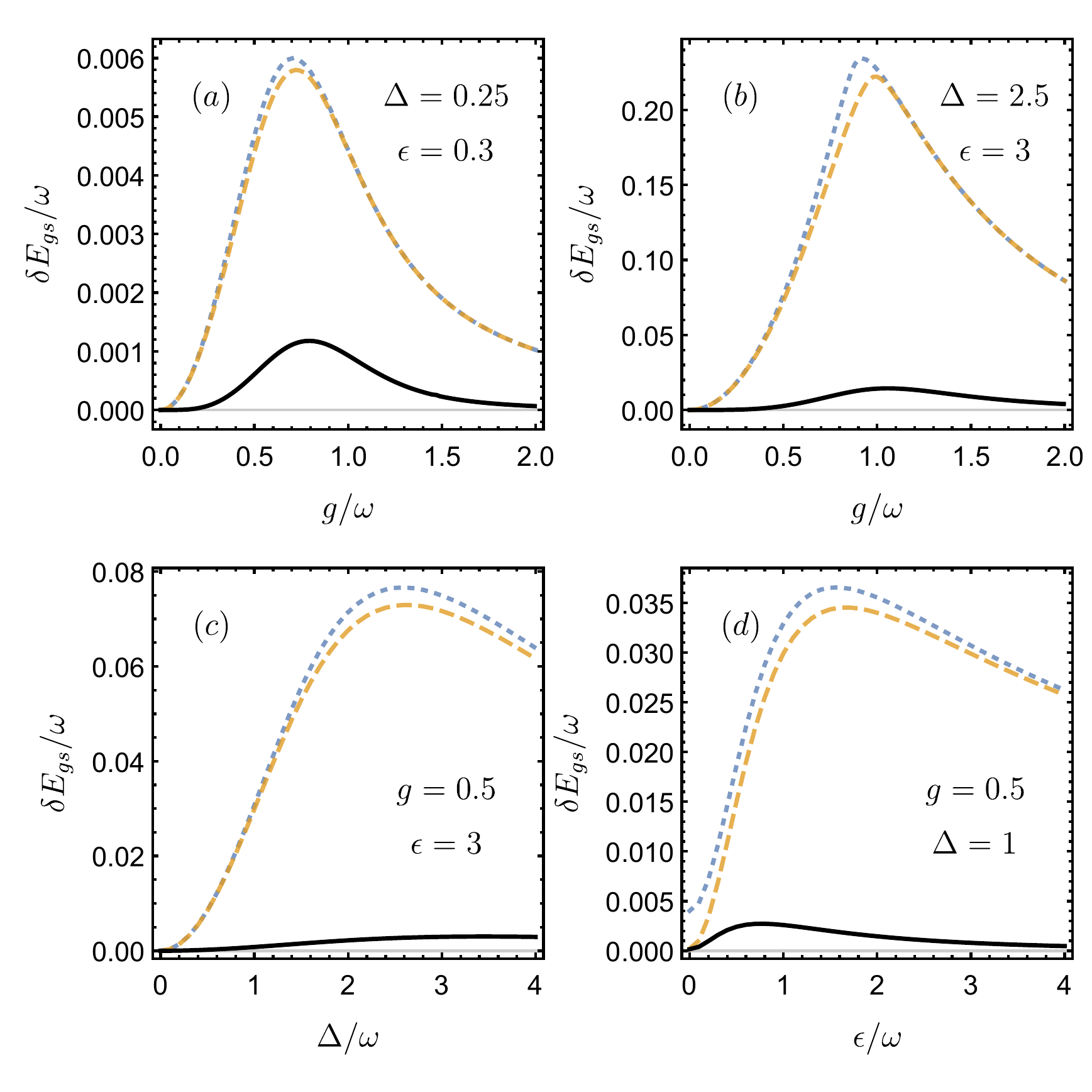}
	}   
	\subfigure{
		\includegraphics[width=.75\linewidth]{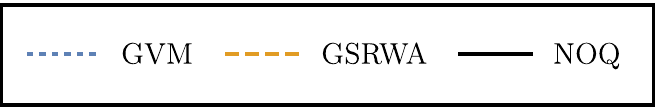}        
	}                          
	\caption{Comparison of the ground state energy deviations for the AQRM obtained using 
	the GVM \cite{Mao_2018}, GSRWA \cite{Xie2020} and the NOQ Ansatz. 
	The parameter values are displayed in the figures, with $\omega=1$.  } 
	\label{EnergyComparison}
\end{figure}

\begin{figure}[t]\centering 
	\subfigure{                
		\centering                                                 
		\includegraphics[width=\linewidth]{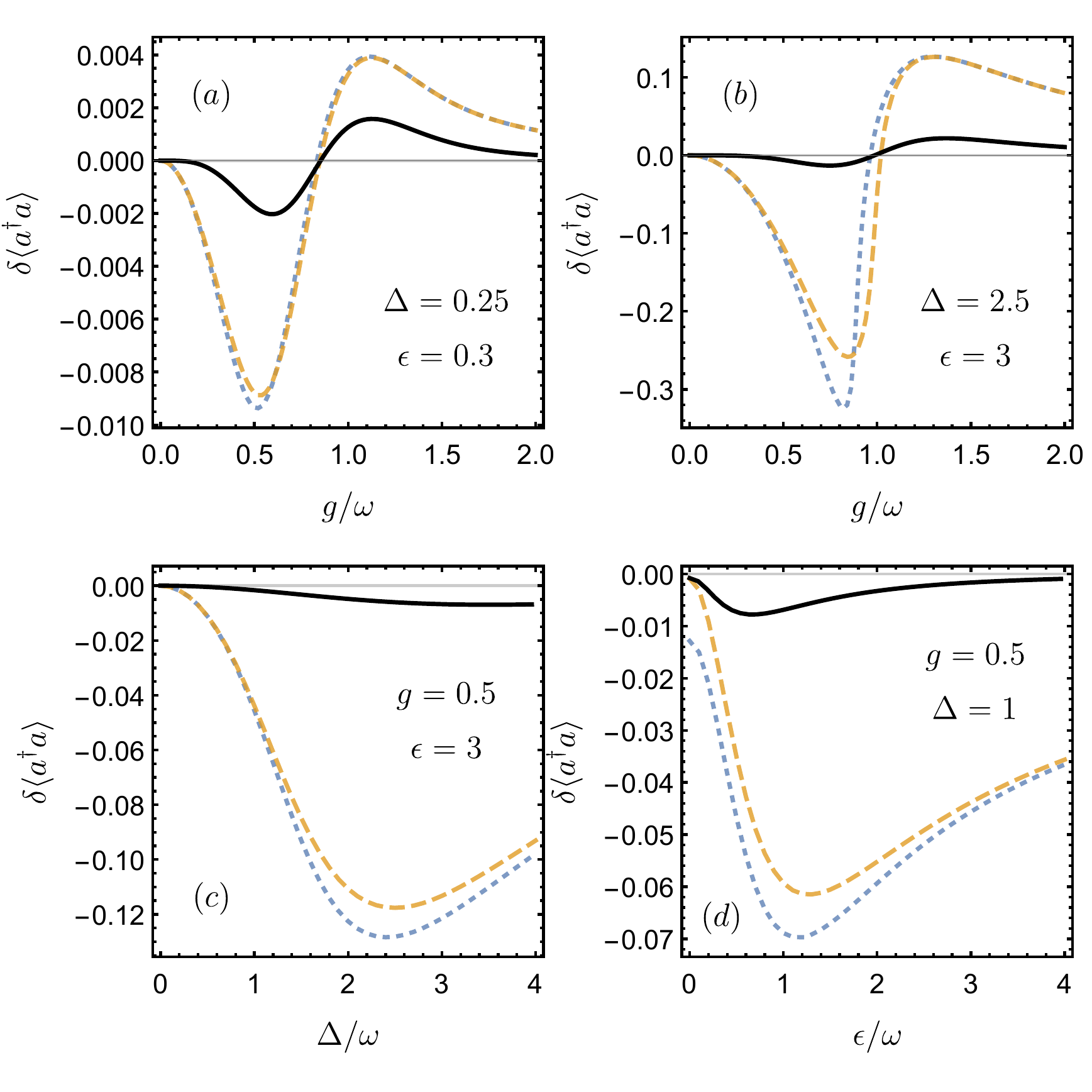}        
	}             
	\subfigure{
		\includegraphics[width=.75\linewidth]{legends.pdf}        
	}                          
	\caption{Comparison of the results obtained for the ground state mean photon number deviations using  
	the same methods as in Fig.~\ref{EnergyComparison}. 
	The parameter values are displayed in the figures, with $\omega=1$.  } 
	\label{PhotonComparison}
\end{figure}

To calculate the ground state energy and other physical observables of interest, we first collect the relevant inner products.
These are
\begin{equation}\label{Inner}
\begin{aligned}
\langle \alpha_i | \alpha_j \rangle &= \delta_{ij} + (1-\delta_{ij}) \, e^{-2\alpha^2},   \\
\langle \phi_i | \phi_j \rangle &= \delta_{ij} + (1-\delta_{ij})\cos\theta,
\end{aligned}
\end{equation}
with $i,j = \pm$. 
The normalization factor follows as
\begin{equation}\label{Normailzation}
\mathcal{N}=\langle \psi_0 | \psi_0 \rangle = 1-2p\sqrt{1-p^2} \, e^{-2\alpha^2}\cos\theta. 
\end{equation}
The ground state energy consists of some physical observables of interest, namely
\begin{equation}\label{Egs}
E_{gs}  =\dfrac{\Delta}{2}\langle \sigma_z\rangle + \omega \langle a^\dagger a \rangle + g\langle \sigma_x \left( a^\dagger + a \right) \rangle  + \dfrac{\epsilon}{2} \langle \sigma_x \rangle,
\end{equation}
where the atomic population is
\begin{equation}\label{Atom}
\langle \sigma_z\rangle = \dfrac{\mathcal{N}-\sin^2\theta}{\mathcal{N}\cos\theta},
\end{equation}
the mean photon number is
\begin{equation}\label{Photon}
\langle a^\dagger a \rangle =  \alpha^2\left(\dfrac{2}{\mathcal{N}} - 1\right),
\end{equation}
the qubit-photon correlation is
\begin{equation}\label{Correlation}
\langle \sigma_x \left( a^\dagger + a \right) \rangle = \dfrac{-2\alpha\sin\theta}{\mathcal{N}},
\end{equation}
and the qubit orientation in the $x$-direction is
\begin{equation}\label{Orientation}
\langle \sigma_x \rangle = \dfrac{(1-2p^2)\sin\theta}{\mathcal{N}}.
\end{equation}
To incorporate squeezing effects if required, 
it is relatively straightforward to show that the mean photon number is
\begin{equation}
\langle a^\dagger a \rangle_{\gamma} =  \alpha^2\left(\dfrac{2}{\mathcal{N}}\cosh 2\gamma - e^{2\gamma}\right) + \sinh^2 \gamma,
\end{equation}
and the qubit-photon correlation is
\begin{equation}
\langle \sigma_x \left( a^\dagger + a \right) \rangle_{\gamma} = \dfrac{-2\alpha\sin\theta}{\mathcal{N}} e^{-\gamma}.
\end{equation}
The atomic population and the qubit orientation in the $x$-direction remain unchanged.

According to the standard variational approach, we need to solve the equations 
\begin{equation}\label{Gradient}
\dfrac{\partial E_{gs}}{\partial \alpha} =\dfrac{\partial E_{gs}}{\partial \theta} =\dfrac{\partial E_{gs}}{\partial p} =0.
\end{equation}
These are transcendental equations which cannot be solved easily through analytic methods. 
Nevertheless, they are readily solved numerically \footnote{For example, in \textit{Mathematica 12} we simply use the function \textbf{NMinimize} 
to minimize the ground state energy defined in Eq.~(\ref{Egs})}.

\begin{figure}[t]\centering  
	\subfigure{                
		\centering                                                 
		\includegraphics[width=\linewidth]{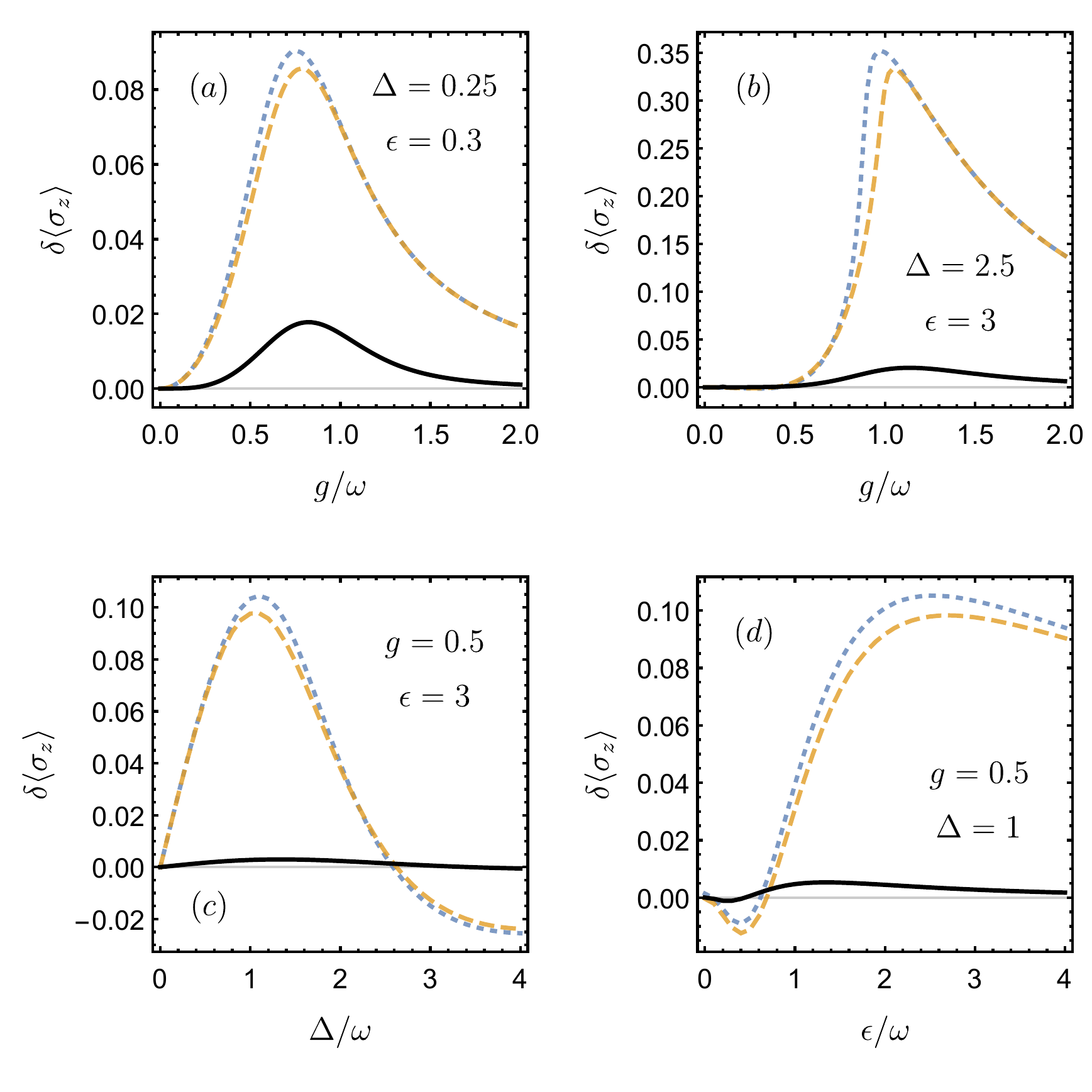}        
	}  
	\subfigure{
		\includegraphics[width=.75\linewidth]{legends.pdf}        
	}                          
	\caption{Comparison of the results obtained for the atomic population $\langle \sigma_z \rangle$ deviations 
	in the ground state using the same methods 
	as in Fig.~\ref{EnergyComparison}. The parameter values are displayed in the figures, with $\omega=1$.  } 
	\label{SigmaZComparison}
\end{figure}

\begin{figure}[t]\centering  
	\subfigure{                
		\centering                                                 
		\includegraphics[width=\linewidth]{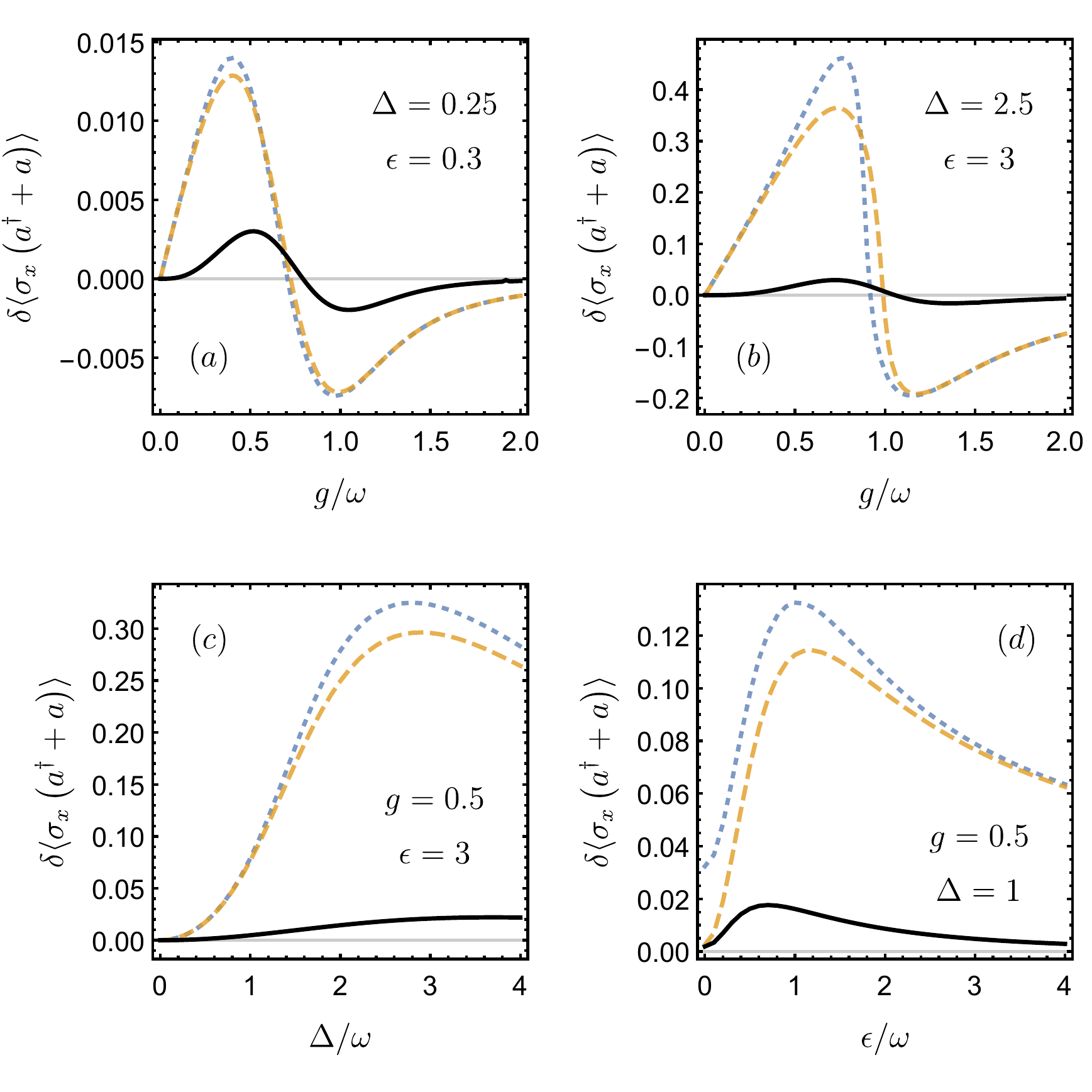}        
	}   
	\subfigure{
		\includegraphics[width=.75\linewidth]{legends.pdf}        
	}                          
	\caption{Comparison of the results obtained for the qubit-cavity correlation $\langle \sigma_x \left(a^\dagger + a \right) \rangle$ deviations 
	in the ground state using the same methods as in Fig.~\ref{EnergyComparison}. 
	The parameter values are displayed in the figures, with $\omega=1$.  } 
	\label{CorrelationComparison}
\end{figure}

In Fig.~\ref{EnergyComparison} we compare the deviation of the ground state energies obtained using the 
approximate methods from the exact results from numerical diagonalization in the truncated Hilbert space. 
In this way we compare the results obtained using the NOQ Ansatz with 
the GVM and GSRWA approximation schemes \cite{Zhang_2013,Mao_2018,Xie2020}.
Even though the authors of the GVM \cite{Mao_2018} and the GSRWA \cite{Xie2020} obtained approximate analytic expressions for some specific parameter regimes, we still calculate eigenvalues and other physical observables variationally, 
in order to make the best comparison with their trial functions. 
For relatively small $\epsilon$, the previous approximations work reasonably well in the coupling regimes where $g/\omega\le 1$.
However, for large $\epsilon$, all previous approximations deviate substantially from the exact results. 
In contrast, the NOQ Ansatz performs quite well in all parameter regimes. 
Only very small deviations are observed around $g=1$. 
As mentioned earlier, with extremely large $\Delta/\omega$, we need to take the squeezing effect into consideration. 
In the regimes where $\Delta/\omega<5$, however, optimizing the squeezing parameter does not make any noticeable difference to the eigenvalues.

Apart from the eigenvalues of the ground state, we also calculate the mean photon number, shown in Fig.~\ref{PhotonComparison}, 
the atomic population, shown in Fig.~\ref{SigmaZComparison}, and the qubit-cavity correlation, shown in Fig.~\ref{CorrelationComparison}.
These results are based on Eqs.~(\ref{Atom})-(\ref{Correlation}). 
We again make comparisons with the previous approximations in terms of their deviation from the exact numerical results.
The parameters are chosen to cover a wide range of regimes. 
Again, the NOQ Ansatz is seen to agree excellently with the exact results in all cases.

We stress that the near perfect agreement is not limited to the displayed regimes, but also with arbitrary parameter values. 
The accuracy of Eq.~(\ref{NOQ}) proves that the NOQ Ansatz has indeed captured the nature of the light-matter interaction process in the AQRM. 
Although, as already mentioned, for very large $\Delta/\omega$, the squeezing effect can no longer be simply omitted.

\section{Discussion}\label{SectionDiscussion}

\subsection{Validity analysis}
\begin{figure}[htbp]
	\subfigure{
		\includegraphics[width=\linewidth]{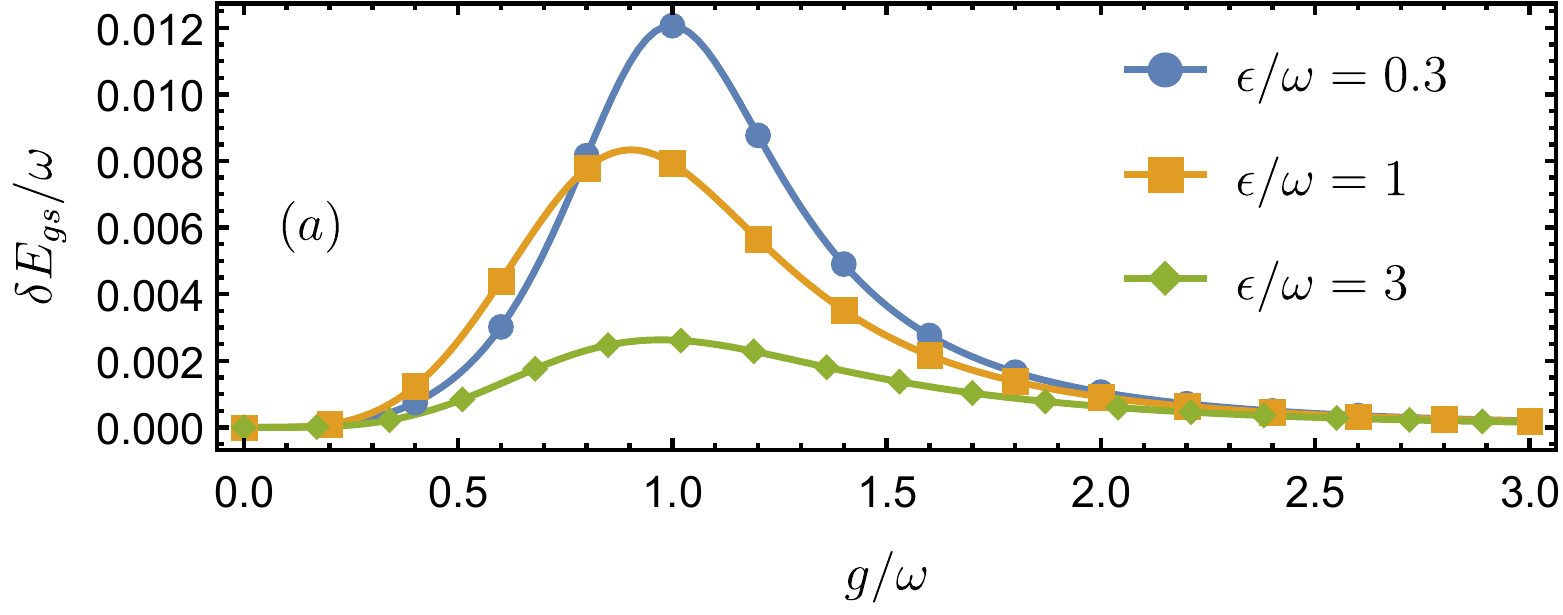}
		\label{Deviation}
	}
	\subfigure{
		\includegraphics[width=\linewidth]{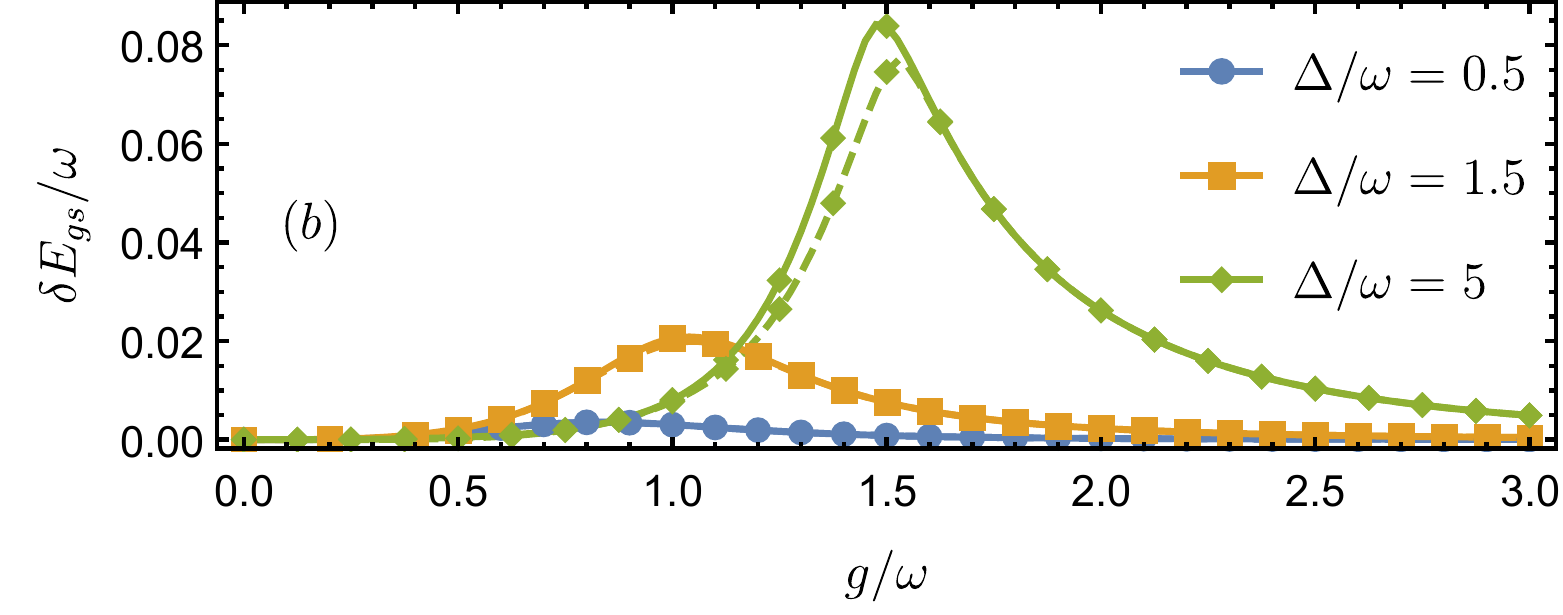}
		\label{Deviation2}
	}
	\subfigure{
		\includegraphics[width=\linewidth]{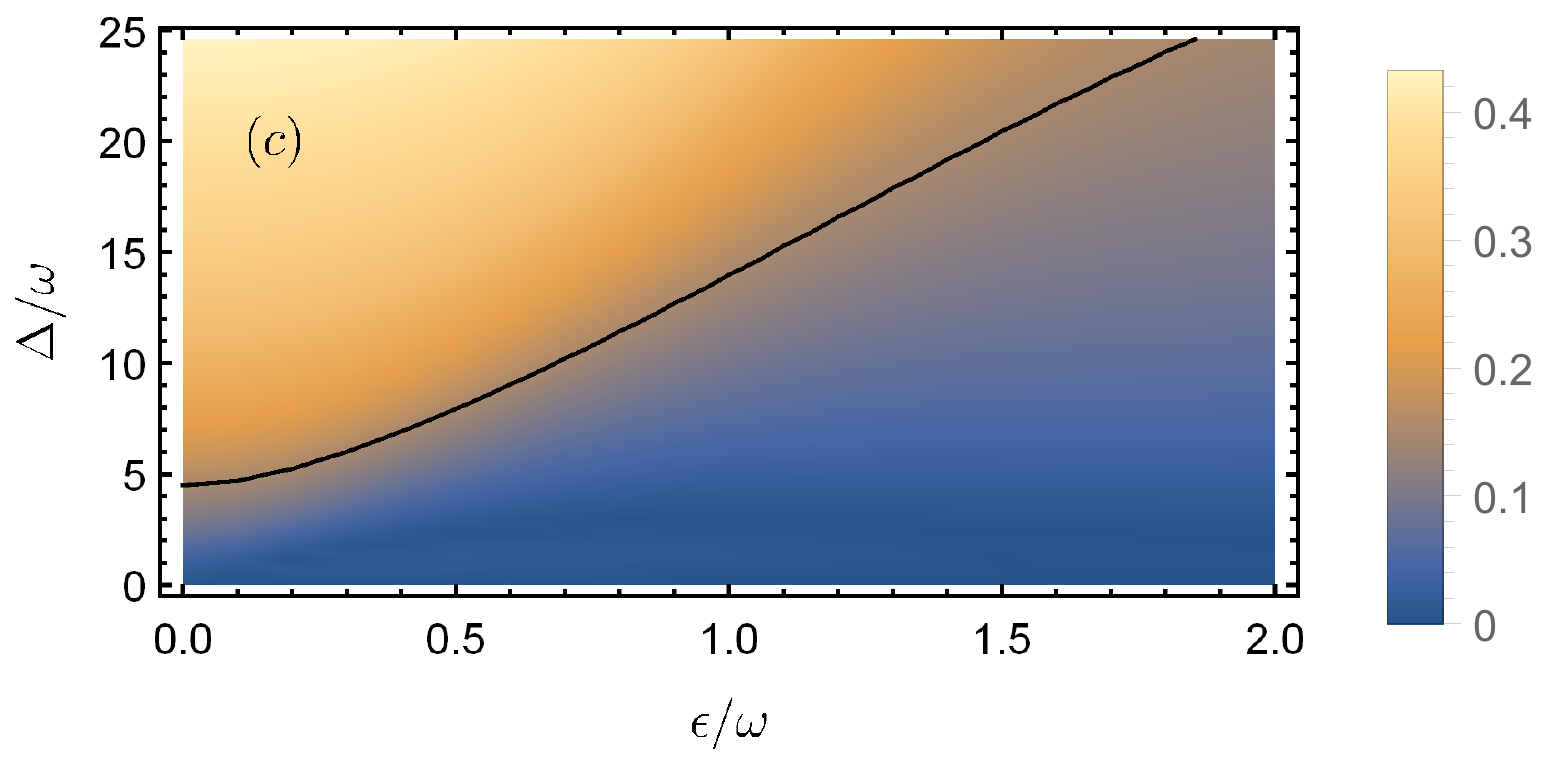}
		\label{gammaDensity}
	}
	\caption{(a) Difference between the energy determined through Eq.~(\ref{Egs}) and the exact results with fixed $\Delta/\omega=1$ and various values of $\epsilon$. Slight deviation around $g/\omega=1$ is observed. (b) Energy deviations with fixed $\epsilon/\omega=0.5$ and various values of $\Delta$. Dashed lines are calculated with the squeezing effect included.  (c) Absolute value of optimal squeezing parameter $|\gamma|$ with respect to different values of $\Delta$ and $\epsilon$. The contour line of $|\gamma|=0.1$ is indicated. The coupling strength is determined by $g=\sqrt{\Delta\omega}/2$, corresponding to the critical coupling strength for the superradiant phase transition in the QRM.}
\end{figure}

We have demonstrated the accuracy of the NOQ Ansatz in almost all parameter regimes.
Nevertheless, it is still meaningful to further discuss the validity of the approximation. 
In most cases the NOQ Ansatz precisely describes the ground state eigenvalue and other static physical properties of the AQRM ground state, 
with however, a tiny deviation from the exact results observed around $g/\omega=1$ in Fig.~\ref{EnergyComparison}. 
This deviation is more obvious if we focus on the difference between the energy determined from Eq.~(\ref{Egs}) and the exact result, as shown in 
Figs. \ref{Deviation} and \ref{Deviation2}. 
It can be seen that the small deviation occurs in intermediate coupling regimes for different parameter values. 
It is also apparent from these figures that our Ansatz is more accurate with larger $\epsilon$ and smaller $\Delta$. 
The reasons for the small deviation may originate in the number of variational parameters. 
Firstly, the physical motivation given in section \ref{SectionNOQ} simplifies some facts.
For instance, the displacements associated with two different qubit states are not necessarily the same. 
On the other hand, the non-orthogonal qubit states may not always have equal rotating angles. 
One could further improve the accuracy by imposing more variational parameters \cite{Hwang_2010,Chen_2020}. 
However, having too complicated a trial function may not only increase the computational difficulties, but also will conceal the underlying physics. 
Another possible reason for the deviation is the complicated multi-photon processes in this coupling regime \cite{Mao_2018}. 

Now we consider the necessity of the squeezing effect for large $\Delta$. 
Intuitively, the effective potentials in Fig.~\ref{EffectivePotentials} are harmonic when there is no tunnelling. 
In the presence of tunnelling, the potentials become non-trivially anharmonic, and the oscillator frequencies are changed accordingly. 
The modulation of frequency can be described by the squeezing effect in phase space \cite{Fox2006}. 
As we have seen in the previous section, this modulation is negligible for the parameter regimes of interest, i.e., with $\Delta/\omega\lesssim 1$. 
However, the tunnelling-induced anharmonicity dominates when the tunnelling parameter $\Delta$ becomes large. 
When $\Delta/\omega\rightarrow \infty$ and $\epsilon=0$, a superradiance phase transition occurs. 
The squeezing effect is strongest around the critical coupling point determined by $g = {\sqrt{\Delta\omega}}/{2}$ \cite{Hwang_2010,Chen_2020}.
In this case, omitting the squeezing effect may yield inaccurate results. 
The absolute value of the optimal squeezing parameter $|\gamma|$ with respect to different values of $\Delta$ and $\epsilon$ is displayed in 
Fig.~\ref{gammaDensity}. 
Also shown is the contour line of $|\gamma|=0.1$.  
Including the squeezing parameter brings noticeable improvement when $|\gamma|\gtrsim 0.1$. 
Examples are displayed with dashed lines in Fig.~\ref{Deviation2}. 
The squeezing parameter lowers the energy deviation of the $\Delta/\omega=5$ case in the intermediate regime, 
whereas the improvement for smaller $\Delta$ values is negligible. 
It is clear that in the parameter regimes of interest, $\gamma$ can be safely dropped.

\subsection{Relation to other approximations}

\begin{figure}[htb]
	\includegraphics[width=.9\linewidth]{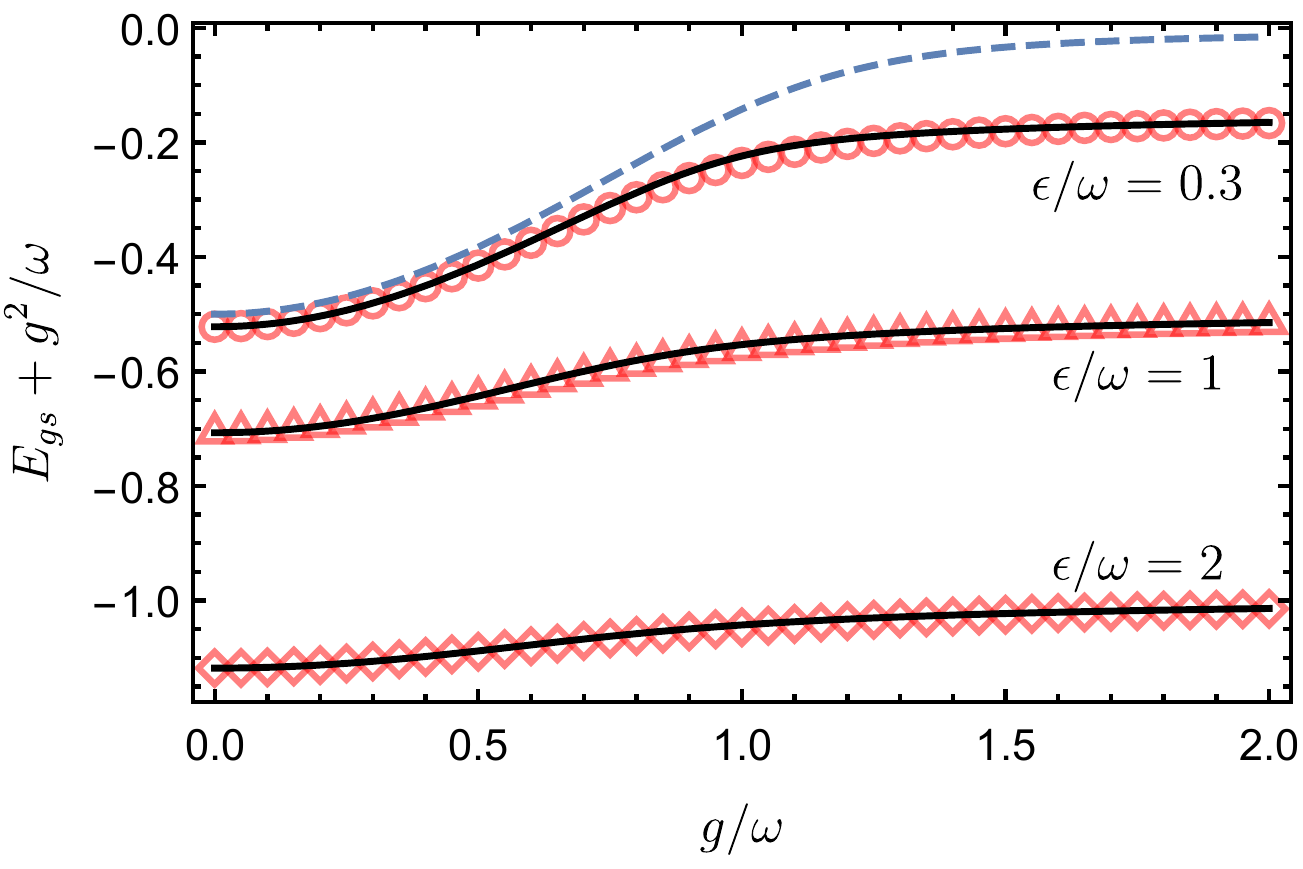}
	\caption{Ground state energy of the AQRM with $\Delta=1, \omega=1$ and values of $\epsilon$ as indicated. 
	The results obtained from our variational Ansatz are denoted by solid lines, with the exact results denoted by open markers. 
	The dashed line is the result obtained with the fixed weight parameter $p={1}/{\sqrt{2}}$, 
	which only works for $\epsilon=0$ and does not capture changes in $\epsilon$.  }
	\label{ComparisonWithNOQforQRM}
\end{figure}

The non-orthogonal qubit states approach was exploited to study the standard QRM \cite{Irish_2014,Leroux_2017}, where parity plays a crucial role. 
Here in the AQRM, however, parity symmetry is broken by the asymmetric term with nonzero $\epsilon$, leading to unequal contributions from different effective potentials, as displayed in Fig.~\ref{EffectivePotentialsAQRM}. 
Meanwhile, the ground state energy is lowered by nonzero $\epsilon$. 
This breaking of symmetry is encoded in the variational parameter $p$ appearing in Eq.~(\ref{NOQ}). 
Without this weight parameter $p$, the wavefunction is independent of $\epsilon$, as can be seen in 
Fig.~\ref{ComparisonWithNOQforQRM}. 
When $\epsilon=0$, the AQRM reduces to the symmetric QRM, and the parameter $p$ in Eq.~(\ref{NOQ}) 
becomes the constant ${1}/{\sqrt{2}}$ as expected, coinciding with the variational Ansatz and 
results for the QRM in Refs.~\cite{Irish_2014,Leroux_2017}. 

We also notice that there is a correspondence between the NOQ Ansatz and the so-called polaron picture \cite{Irish_2014,Ying_2015,Liu2017a,Cong_2017,*Cong_2019,Sun2019}. 
This correspondence is not immediately obvious in the form of Eq.~(\ref{NOQ}). 
To see this, we expand the non-orthogonal qubit states $| \phi_\pm\rangle$ and write the Ansatz in the $| \! \pm z\rangle$ basis as
\begin{equation}\label{NOQzbasis}
\begin{aligned}
\psi_0^z = & \cos\dfrac{\theta}{2}\left(p \, |\alpha\rangle-\sqrt{1-p^2} \, |-\alpha\rangle\right) \, |\!+z\rangle  \\
& - \sin\dfrac{\theta}{2}\left(p \, |\alpha\rangle+\sqrt{1-p^2} \, |-\alpha\rangle\right) \, |\!-z\rangle.
\end{aligned}
\end{equation}
Now the cavity states $|\!\pm\alpha\rangle$ correspond to the polaron and anti-polaron associated with the qubit states $|\!\pm z \rangle$.

\section{Conclusion}\label{SectionConclusion}

In conclusion, starting from physical observations, we have proposed an Ansatz to describe the ground state of the AQRM, 
based on a weighted superposition of squeezed coherent states and non-orthogonal qubit (NOQ) states. 
The Ansatz relies only on three variational parameters ($\alpha,\theta,p$), and precisely describes the system ground state, 
exhibiting very good agreement with the numerical exact solution.
A fourth variational parameter can also be introduced based on the squeezing parameter, 
but does not make any noticeable difference to the eigenvalues.
As shown in Figs~\ref{EnergyComparison}-\ref{CorrelationComparison} for the ground state energy, the ground state mean photon number, 
the atomic population in the ground state, and the qubit cavity correlation in the  ground state, 
the Ansatz, with the three variational parameters, works remarkably well in almost all parameter regimes.
These figures also show a comparison with the existing approximations \cite{Zhang_2013,Mao_2018,Xie2020}, 
and clear qualitative improvements are observed.

The asymmetric term in the AQRM arises naturally in the cQED setup, 
and can be conveniently manipulated externally. 
It is expected that the asymmetric counterparts of other currently studied light-matter interaction models, such as the anisotropic QRM \cite{Tomka_2014,Xie_2014} and the Rabi-Stark model \cite{Eckle_2017,Xie_2019,Cong_2020,Xie_2020} will be realized in future experiments. 
The NOQ Ansatz presented here can be straightforwardly generalized to these related models for which similarly good performance is expected.

We conclude by emphasizing that the generalization of the NOQ Ansatz to excited states remains an open question. 
It would be particularly worthwhile, with regard to investigating topological properties, 
if the conical intersection structure \cite{Batchelor2015} in the energy spectrum of the AQRM 
can be recovered correctly in this way. 

~

{\bf{Acknowledgments}}. The authors are grateful to David Tilbrook for illuminating discussions on cQED. 
This work is supported by the Australian Research Council grant DP170104934 and the 
 National Natural Science Foundation of China (Grant No. 11174375). 


\begin{thebibliography}{65}%
	\makeatletter
	\providecommand \@ifxundefined [1]{%
		\@ifx{#1\undefined}
	}%
	\providecommand \@ifnum [1]{%
		\ifnum #1\expandafter \@firstoftwo
		\else \expandafter \@secondoftwo
		\fi
	}%
	\providecommand \@ifx [1]{%
		\ifx #1\expandafter \@firstoftwo
		\else \expandafter \@secondoftwo
		\fi
	}%
	\providecommand \natexlab [1]{#1}%
	\providecommand \enquote  [1]{``#1''}%
	\providecommand \bibnamefont  [1]{#1}%
	\providecommand \bibfnamefont [1]{#1}%
	\providecommand \citenamefont [1]{#1}%
	\providecommand \href@noop [0]{\@secondoftwo}%
	\providecommand \href [0]{\begingroup \@sanitize@url \@href}%
	\providecommand \@href[1]{\@@startlink{#1}\@@href}%
	\providecommand \@@href[1]{\endgroup#1\@@endlink}%
	\providecommand \@sanitize@url [0]{\catcode `\\12\catcode `\$12\catcode
		`\&12\catcode `\#12\catcode `\^12\catcode `\_12\catcode `\%12\relax}%
	\providecommand \@@startlink[1]{}%
	\providecommand \@@endlink[0]{}%
	\providecommand \url  [0]{\begingroup\@sanitize@url \@url }%
	\providecommand \@url [1]{\endgroup\@href {#1}{\urlprefix }}%
	\providecommand \urlprefix  [0]{URL }%
	\providecommand \Eprint [0]{\href }%
	\providecommand \doibase [0]{http://dx.doi.org/}%
	\providecommand \selectlanguage [0]{\@gobble}%
	\providecommand \bibinfo  [0]{\@secondoftwo}%
	\providecommand \bibfield  [0]{\@secondoftwo}%
	\providecommand \translation [1]{[#1]}%
	\providecommand \BibitemOpen [0]{}%
	\providecommand \bibitemStop [0]{}%
	\providecommand \bibitemNoStop [0]{.\EOS\space}%
	\providecommand \EOS [0]{\spacefactor3000\relax}%
	\providecommand \BibitemShut  [1]{\csname bibitem#1\endcsname}%
	\let\auto@bib@innerbib\@empty
	\bibitem [{\citenamefont {Rabi}(1936)}]{Rabi_1936}%
	\BibitemOpen
	\bibfield  {author} {\bibinfo {author} {\bibfnamefont {I.~I.}\ \bibnamefont
			{Rabi}},\ }\href {\doibase 10.1103/physrev.49.324} {\bibfield  {journal}
		{\bibinfo  {journal} {Phys. Rev.}\ }\textbf {\bibinfo {volume} {49}},\
		\bibinfo {pages} {324} (\bibinfo {year} {1936})}\BibitemShut {NoStop}%
	\bibitem [{\citenamefont {Rabi}(1937)}]{Rabi_1937}%
	\BibitemOpen
	\bibfield  {author} {\bibinfo {author} {\bibfnamefont {I.~I.}\ \bibnamefont
			{Rabi}},\ }\href {\doibase 10.1103/physrev.51.652} {\bibfield  {journal}
		{\bibinfo  {journal} {Phys. Rev.}\ }\textbf {\bibinfo {volume} {51}},\
		\bibinfo {pages} {652} (\bibinfo {year} {1937})}\BibitemShut {NoStop}%
	\bibitem [{\citenamefont {Scully}\ and\ \citenamefont
		{Zubairy}(1997)}]{Scully_1997}%
	\BibitemOpen
	\bibfield  {author} {\bibinfo {author} {\bibfnamefont {M.~O.}\ \bibnamefont
			{Scully}}\ and\ \bibinfo {author} {\bibfnamefont {M.~S.}\ \bibnamefont
			{Zubairy}},\ }\href {\doibase 10.1017/cbo9780511813993} {\emph {\bibinfo
			{title} {Quantum Optics}}}\ (\bibinfo  {publisher} {Cambridge University
		Press},\ \bibinfo {year} {1997})\BibitemShut {NoStop}%
	\bibitem [{\citenamefont {Fox}(2006)}]{Fox2006}%
	\BibitemOpen
	\bibfield  {author} {\bibinfo {author} {\bibfnamefont {M.}~\bibnamefont
			{Fox}},\ }\href
	{https://www.ebook.de/de/product/5214114/mark_fox_quantum_optics.html} {\emph
		{\bibinfo {title} {Quantum Optics}}}\ (\bibinfo  {publisher} {Oxford
		University Press},\ \bibinfo {year} {2006})\BibitemShut {NoStop}%
	\bibitem [{\citenamefont {Wagner}(1986)}]{Wagner1986}%
	\BibitemOpen
	\bibfield  {author} {\bibinfo {author} {\bibfnamefont {M.}~\bibnamefont
			{Wagner}},\ }\href@noop {} {\emph {\bibinfo {title} {Unitary transformations
				in solid state physics}}}\ (\bibinfo  {publisher} {North-Holland},\ \bibinfo
	{address} {Amsterdam},\ \bibinfo {year} {1986})\BibitemShut {NoStop}%
	\bibitem [{\citenamefont {Braak}\ \emph {et~al.}(2016)\citenamefont {Braak},
		\citenamefont {Chen}, \citenamefont {Batchelor},\ and\ \citenamefont
		{Solano}}]{Braak_2016}%
	\BibitemOpen
	\bibfield  {author} {\bibinfo {author} {\bibfnamefont {D.}~\bibnamefont
			{Braak}}, \bibinfo {author} {\bibfnamefont {Q.-H.}\ \bibnamefont {Chen}},
		\bibinfo {author} {\bibfnamefont {M.~T.}\ \bibnamefont {Batchelor}}, \ and\
		\bibinfo {author} {\bibfnamefont {E.}~\bibnamefont {Solano}},\ }\href
	{\doibase 10.1088/1751-8113/49/30/300301} {\bibfield  {journal} {\bibinfo
			{journal} {J. Phys. A: Math. Theor.}\ }\textbf {\bibinfo {volume} {49}},\
		\bibinfo {pages} {300301} (\bibinfo {year} {2016})}\BibitemShut {NoStop}%
	\bibitem [{\citenamefont {Wendin}(2017)}]{Wendin_2017}%
	\BibitemOpen
	\bibfield  {author} {\bibinfo {author} {\bibfnamefont {G.}~\bibnamefont
			{Wendin}},\ }\href {\doibase 10.1088/1361-6633/aa7e1a} {\bibfield  {journal}
		{\bibinfo  {journal} {Rep. Prog. Phys.}\ }\textbf {\bibinfo {volume} {80}},\
		\bibinfo {pages} {106001} (\bibinfo {year} {2017})}\BibitemShut {NoStop}%
	\bibitem [{\citenamefont {Forn-D\'{\i}az}\ \emph {et~al.}(2019)\citenamefont
		{Forn-D\'{\i}az}, \citenamefont {Lamata}, \citenamefont {Rico}, \citenamefont
		{Kono},\ and\ \citenamefont {Solano}}]{Forn_D_az_2019}%
	\BibitemOpen
	\bibfield  {author} {\bibinfo {author} {\bibfnamefont {P.}~\bibnamefont
			{Forn-D\'{\i}az}}, \bibinfo {author} {\bibfnamefont {L.}~\bibnamefont
			{Lamata}}, \bibinfo {author} {\bibfnamefont {E.}~\bibnamefont {Rico}},
		\bibinfo {author} {\bibfnamefont {J.}~\bibnamefont {Kono}}, \ and\ \bibinfo
		{author} {\bibfnamefont {E.}~\bibnamefont {Solano}},\ }\href {\doibase
		10.1103/RevModPhys.91.025005} {\bibfield  {journal} {\bibinfo  {journal}
			{Rev. Mod. Phys.}\ }\textbf {\bibinfo {volume} {91}},\ \bibinfo {pages}
		{025005} (\bibinfo {year} {2019})}\BibitemShut {NoStop}%
	\bibitem [{\citenamefont {Kockum}\ \emph {et~al.}(2019)\citenamefont {Kockum},
		\citenamefont {Miranowicz}, \citenamefont {Liberato}, \citenamefont
		{Savasta},\ and\ \citenamefont {Nori}}]{Frisk_Kockum_2019}%
	\BibitemOpen
	\bibfield  {author} {\bibinfo {author} {\bibfnamefont {A.~F.}\ \bibnamefont
			{Kockum}}, \bibinfo {author} {\bibfnamefont {A.}~\bibnamefont {Miranowicz}},
		\bibinfo {author} {\bibfnamefont {S.~D.}\ \bibnamefont {Liberato}}, \bibinfo
		{author} {\bibfnamefont {S.}~\bibnamefont {Savasta}}, \ and\ \bibinfo
		{author} {\bibfnamefont {F.}~\bibnamefont {Nori}},\ }\href {\doibase
		10.1038/s42254-018-0006-2} {\bibfield  {journal} {\bibinfo  {journal} {Nat.
				Rev. Phys.}\ }\textbf {\bibinfo {volume} {1}},\ \bibinfo {pages} {19}
		(\bibinfo {year} {2019})}\BibitemShut {NoStop}%
	\bibitem [{\citenamefont {Blais}\ \emph {et~al.}()\citenamefont {Blais},
		\citenamefont {Grimsmo}, \citenamefont {Girvin},\ and\ \citenamefont
		{Wallraff}}]{Blais2020}%
	\BibitemOpen
	\bibfield  {author} {\bibinfo {author} {\bibfnamefont {A.}~\bibnamefont
			{Blais}}, \bibinfo {author} {\bibfnamefont {A.~L.}\ \bibnamefont {Grimsmo}},
		\bibinfo {author} {\bibfnamefont {S.~M.}\ \bibnamefont {Girvin}}, \ and\
		\bibinfo {author} {\bibfnamefont {A.}~\bibnamefont {Wallraff}},\ }\href@noop
	{} {\ }\Eprint {http://arxiv.org/abs/2005.12667} {arXiv:2005.12667}
	\BibitemShut {NoStop}%
	\bibitem [{\citenamefont {Braak}(2011)}]{Braak_2011}%
	\BibitemOpen
	\bibfield  {author} {\bibinfo {author} {\bibfnamefont {D.}~\bibnamefont
			{Braak}},\ }\href {\doibase 10.1103/PhysRevLett.107.100401} {\bibfield
		{journal} {\bibinfo  {journal} {Phys. Rev. Lett.}\ }\textbf {\bibinfo
			{volume} {107}},\ \bibinfo {pages} {100401} (\bibinfo {year}
		{2011})}\BibitemShut {NoStop}%
	\bibitem [{\citenamefont {Chen}\ \emph {et~al.}(2012)\citenamefont {Chen},
		\citenamefont {Wang}, \citenamefont {He}, \citenamefont {Liu},\ and\
		\citenamefont {Wang}}]{Chen2012}%
	\BibitemOpen
	\bibfield  {author} {\bibinfo {author} {\bibfnamefont {Q.-H.}\ \bibnamefont
			{Chen}}, \bibinfo {author} {\bibfnamefont {C.}~\bibnamefont {Wang}}, \bibinfo
		{author} {\bibfnamefont {S.}~\bibnamefont {He}}, \bibinfo {author}
		{\bibfnamefont {T.}~\bibnamefont {Liu}}, \ and\ \bibinfo {author}
		{\bibfnamefont {K.-L.}\ \bibnamefont {Wang}},\ }\href {\doibase
		10.1103/PhysRevA.86.023822} {\bibfield  {journal} {\bibinfo  {journal} {Phys.
				Rev. A}\ }\textbf {\bibinfo {volume} {86}},\ \bibinfo {pages} {023822}
		(\bibinfo {year} {2012})}\BibitemShut {NoStop}%
	\bibitem [{\citenamefont {Zhong}\ \emph {et~al.}(2013)\citenamefont {Zhong},
		\citenamefont {Xie}, \citenamefont {Batchelor},\ and\ \citenamefont
		{Lee}}]{Zhong_2013}%
	\BibitemOpen
	\bibfield  {author} {\bibinfo {author} {\bibfnamefont {H.}~\bibnamefont
			{Zhong}}, \bibinfo {author} {\bibfnamefont {Q.}~\bibnamefont {Xie}}, \bibinfo
		{author} {\bibfnamefont {M.~T.}\ \bibnamefont {Batchelor}}, \ and\ \bibinfo
		{author} {\bibfnamefont {C.}~\bibnamefont {Lee}},\ }\href {\doibase
		10.1088/1751-8113/46/41/415302} {\bibfield  {journal} {\bibinfo  {journal}
			{J. Phys. A: Math. Theor.}\ }\textbf {\bibinfo {volume} {46}},\ \bibinfo
		{pages} {415302} (\bibinfo {year} {2013})}\BibitemShut {NoStop}%
	\bibitem [{\citenamefont {Zhong}\ \emph {et~al.}(2014)\citenamefont {Zhong},
		\citenamefont {Xie}, \citenamefont {Guan}, \citenamefont {Batchelor},
		\citenamefont {Gao},\ and\ \citenamefont {Lee}}]{Zhong_2014}%
	\BibitemOpen
	\bibfield  {author} {\bibinfo {author} {\bibfnamefont {H.}~\bibnamefont
			{Zhong}}, \bibinfo {author} {\bibfnamefont {Q.}~\bibnamefont {Xie}}, \bibinfo
		{author} {\bibfnamefont {X.}~\bibnamefont {Guan}}, \bibinfo {author}
		{\bibfnamefont {M.~T.}\ \bibnamefont {Batchelor}}, \bibinfo {author}
		{\bibfnamefont {K.}~\bibnamefont {Gao}}, \ and\ \bibinfo {author}
		{\bibfnamefont {C.}~\bibnamefont {Lee}},\ }\href {\doibase
		10.1088/1751-8113/47/4/045301} {\bibfield  {journal} {\bibinfo  {journal} {J.
				Phys. A: Math. Theor.}\ }\textbf {\bibinfo {volume} {47}},\ \bibinfo {pages}
		{045301} (\bibinfo {year} {2014})}\BibitemShut {NoStop}%
	\bibitem [{\citenamefont {Maciejewski}\ \emph
		{et~al.}(2014{\natexlab{a}})\citenamefont {Maciejewski}, \citenamefont
		{Przybylska},\ and\ \citenamefont {Stachowiak}}]{Maciejewski2014}%
	\BibitemOpen
	\bibfield  {author} {\bibinfo {author} {\bibfnamefont {A.~J.}\ \bibnamefont
			{Maciejewski}}, \bibinfo {author} {\bibfnamefont {M.}~\bibnamefont
			{Przybylska}}, \ and\ \bibinfo {author} {\bibfnamefont {T.}~\bibnamefont
			{Stachowiak}},\ }\href {\doibase 10.1016/j.physleta.2013.10.032} {\bibfield
		{journal} {\bibinfo  {journal} {Phys. Lett. A}\ }\textbf {\bibinfo {volume}
			{378}},\ \bibinfo {pages} {16} (\bibinfo {year}
		{2014}{\natexlab{a}})}\BibitemShut {NoStop}%
	\bibitem [{\citenamefont {Maciejewski}\ \emph
		{et~al.}(2014{\natexlab{b}})\citenamefont {Maciejewski}, \citenamefont
		{Przybylska},\ and\ \citenamefont {Stachowiak}}]{Maciejewski_2014}%
	\BibitemOpen
	\bibfield  {author} {\bibinfo {author} {\bibfnamefont {A.~J.}\ \bibnamefont
			{Maciejewski}}, \bibinfo {author} {\bibfnamefont {M.}~\bibnamefont
			{Przybylska}}, \ and\ \bibinfo {author} {\bibfnamefont {T.}~\bibnamefont
			{Stachowiak}},\ }\href {\doibase 10.1016/j.physleta.2014.10.001} {\bibfield
		{journal} {\bibinfo  {journal} {Phys. Lett. A}\ }\textbf {\bibinfo {volume}
			{378}},\ \bibinfo {pages} {3445} (\bibinfo {year}
		{2014}{\natexlab{b}})}\BibitemShut {NoStop}%
	\bibitem [{\citenamefont {Xie}\ \emph {et~al.}(2017)\citenamefont {Xie},
		\citenamefont {Zhong}, \citenamefont {Batchelor},\ and\ \citenamefont
		{Lee}}]{Xie_2017}%
	\BibitemOpen
	\bibfield  {author} {\bibinfo {author} {\bibfnamefont {Q.}~\bibnamefont
			{Xie}}, \bibinfo {author} {\bibfnamefont {H.}~\bibnamefont {Zhong}}, \bibinfo
		{author} {\bibfnamefont {M.~T.}\ \bibnamefont {Batchelor}}, \ and\ \bibinfo
		{author} {\bibfnamefont {C.}~\bibnamefont {Lee}},\ }\href {\doibase
		10.1088/1751-8121/aa5a65} {\bibfield  {journal} {\bibinfo  {journal} {J.
				Phys. A: Math. Theor.}\ }\textbf {\bibinfo {volume} {50}},\ \bibinfo {pages}
		{113001} (\bibinfo {year} {2017})}\BibitemShut {NoStop}%
	\bibitem [{\citenamefont {Braak}(2019)}]{Braak_2019}%
	\BibitemOpen
	\bibfield  {author} {\bibinfo {author} {\bibfnamefont {D.}~\bibnamefont
			{Braak}},\ }\href {\doibase 10.3390/sym11101259} {\bibfield  {journal}
		{\bibinfo  {journal} {Symmetry}\ }\textbf {\bibinfo {volume} {11}},\ \bibinfo
		{pages} {1259} (\bibinfo {year} {2019})}\BibitemShut {NoStop}%
	\bibitem [{\citenamefont {Jaynes}\ and\ \citenamefont
		{Cummings}(1963)}]{Jaynes_1963}%
	\BibitemOpen
	\bibfield  {author} {\bibinfo {author} {\bibfnamefont {E.}~\bibnamefont
			{Jaynes}}\ and\ \bibinfo {author} {\bibfnamefont {F.}~\bibnamefont
			{Cummings}},\ }\href {\doibase 10.1109/proc.1963.1664} {\bibfield  {journal}
		{\bibinfo  {journal} {Proc. {IEEE}}\ }\textbf {\bibinfo {volume} {51}},\
		\bibinfo {pages} {89} (\bibinfo {year} {1963})}\BibitemShut {NoStop}%
	\bibitem [{\citenamefont {Casanova}\ \emph {et~al.}(2010)\citenamefont
		{Casanova}, \citenamefont {Romero}, \citenamefont {Lizuain}, \citenamefont
		{Garc\'{\i}a-Ripoll},\ and\ \citenamefont {Solano}}]{Casanova_2010}%
	\BibitemOpen
	\bibfield  {author} {\bibinfo {author} {\bibfnamefont {J.}~\bibnamefont
			{Casanova}}, \bibinfo {author} {\bibfnamefont {G.}~\bibnamefont {Romero}},
		\bibinfo {author} {\bibfnamefont {I.}~\bibnamefont {Lizuain}}, \bibinfo
		{author} {\bibfnamefont {J.~J.}\ \bibnamefont {Garc\'{\i}a-Ripoll}}, \ and\
		\bibinfo {author} {\bibfnamefont {E.}~\bibnamefont {Solano}},\ }\href
	{\doibase 10.1103/PhysRevLett.105.263603} {\bibfield  {journal} {\bibinfo
			{journal} {Phys. Rev. Lett.}\ }\textbf {\bibinfo {volume} {105}},\ \bibinfo
		{pages} {263603} (\bibinfo {year} {2010})}\BibitemShut {NoStop}%
	\bibitem [{\citenamefont {Rossatto}\ \emph {et~al.}(2017)\citenamefont
		{Rossatto}, \citenamefont {Villas-B\^oas}, \citenamefont {Sanz},\ and\
		\citenamefont {Solano}}]{Rossatto_2017}%
	\BibitemOpen
	\bibfield  {author} {\bibinfo {author} {\bibfnamefont {D.~Z.}\ \bibnamefont
			{Rossatto}}, \bibinfo {author} {\bibfnamefont {C.~J.}\ \bibnamefont
			{Villas-B\^oas}}, \bibinfo {author} {\bibfnamefont {M.}~\bibnamefont {Sanz}},
		\ and\ \bibinfo {author} {\bibfnamefont {E.}~\bibnamefont {Solano}},\ }\href
	{\doibase 10.1103/physreva.96.013849} {\bibfield  {journal} {\bibinfo
			{journal} {Phys. Rev. A}\ }\textbf {\bibinfo {volume} {96}},\ \bibinfo
		{pages} {013849} (\bibinfo {year} {2017})}\BibitemShut {NoStop}%
	\bibitem [{\citenamefont {Forn-D\'{\i}az}\ \emph {et~al.}(2010)\citenamefont
		{Forn-D\'{\i}az}, \citenamefont {Lisenfeld}, \citenamefont {Marcos},
		\citenamefont {Garc\'{\i}a-Ripoll}, \citenamefont {Solano}, \citenamefont
		{Harmans},\ and\ \citenamefont {Mooij}}]{Forn_D_az_2010}%
	\BibitemOpen
	\bibfield  {author} {\bibinfo {author} {\bibfnamefont {P.}~\bibnamefont
			{Forn-D\'{\i}az}}, \bibinfo {author} {\bibfnamefont {J.}~\bibnamefont
			{Lisenfeld}}, \bibinfo {author} {\bibfnamefont {D.}~\bibnamefont {Marcos}},
		\bibinfo {author} {\bibfnamefont {J.~J.}\ \bibnamefont {Garc\'{\i}a-Ripoll}},
		\bibinfo {author} {\bibfnamefont {E.}~\bibnamefont {Solano}}, \bibinfo
		{author} {\bibfnamefont {C.~J. P.~M.}\ \bibnamefont {Harmans}}, \ and\
		\bibinfo {author} {\bibfnamefont {J.~E.}\ \bibnamefont {Mooij}},\ }\href
	{\doibase 10.1103/PhysRevLett.105.237001} {\bibfield  {journal} {\bibinfo
			{journal} {Phys. Rev. Lett.}\ }\textbf {\bibinfo {volume} {105}},\ \bibinfo
		{pages} {237001} (\bibinfo {year} {2010})}\BibitemShut {NoStop}%
	\bibitem [{\citenamefont {Yoshihara}\ \emph {et~al.}(2016)\citenamefont
		{Yoshihara}, \citenamefont {Fuse}, \citenamefont {Ashhab}, \citenamefont
		{Kakuyanagi}, \citenamefont {Saito},\ and\ \citenamefont
		{Semba}}]{Yoshihara_2016}%
	\BibitemOpen
	\bibfield  {author} {\bibinfo {author} {\bibfnamefont {F.}~\bibnamefont
			{Yoshihara}}, \bibinfo {author} {\bibfnamefont {T.}~\bibnamefont {Fuse}},
		\bibinfo {author} {\bibfnamefont {S.}~\bibnamefont {Ashhab}}, \bibinfo
		{author} {\bibfnamefont {K.}~\bibnamefont {Kakuyanagi}}, \bibinfo {author}
		{\bibfnamefont {S.}~\bibnamefont {Saito}}, \ and\ \bibinfo {author}
		{\bibfnamefont {K.}~\bibnamefont {Semba}},\ }\href {\doibase
		10.1038/nphys3906} {\bibfield  {journal} {\bibinfo  {journal} {Nat. Phys.}\
		}\textbf {\bibinfo {volume} {13}},\ \bibinfo {pages} {44} (\bibinfo {year}
		{2016})}\BibitemShut {NoStop}%
	\bibitem [{\citenamefont {Yoshihara}\ \emph {et~al.}(2018)\citenamefont
		{Yoshihara}, \citenamefont {Fuse}, \citenamefont {Ao}, \citenamefont
		{Ashhab}, \citenamefont {Kakuyanagi}, \citenamefont {Saito}, \citenamefont
		{Aoki}, \citenamefont {Koshino},\ and\ \citenamefont
		{Semba}}]{Yoshihara_2018}%
	\BibitemOpen
	\bibfield  {author} {\bibinfo {author} {\bibfnamefont {F.}~\bibnamefont
			{Yoshihara}}, \bibinfo {author} {\bibfnamefont {T.}~\bibnamefont {Fuse}},
		\bibinfo {author} {\bibfnamefont {Z.}~\bibnamefont {Ao}}, \bibinfo {author}
		{\bibfnamefont {S.}~\bibnamefont {Ashhab}}, \bibinfo {author} {\bibfnamefont
			{K.}~\bibnamefont {Kakuyanagi}}, \bibinfo {author} {\bibfnamefont
			{S.}~\bibnamefont {Saito}}, \bibinfo {author} {\bibfnamefont
			{T.}~\bibnamefont {Aoki}}, \bibinfo {author} {\bibfnamefont {K.}~\bibnamefont
			{Koshino}}, \ and\ \bibinfo {author} {\bibfnamefont {K.}~\bibnamefont
			{Semba}},\ }\href {\doibase 10.1103/PhysRevLett.120.183601} {\bibfield
		{journal} {\bibinfo  {journal} {Phys. Rev. Lett.}\ }\textbf {\bibinfo
			{volume} {120}},\ \bibinfo {pages} {183601} (\bibinfo {year}
		{2018})}\BibitemShut {NoStop}%
	\bibitem [{\citenamefont {Feranchuk}\ \emph {et~al.}(1996)\citenamefont
		{Feranchuk}, \citenamefont {Komarov},\ and\ \citenamefont
		{Ulyanenkov}}]{Feranchuk_1996}%
	\BibitemOpen
	\bibfield  {author} {\bibinfo {author} {\bibfnamefont {I.~D.}\ \bibnamefont
			{Feranchuk}}, \bibinfo {author} {\bibfnamefont {L.~I.}\ \bibnamefont
			{Komarov}}, \ and\ \bibinfo {author} {\bibfnamefont {A.~P.}\ \bibnamefont
			{Ulyanenkov}},\ }\href {\doibase 10.1088/0305-4470/29/14/026} {\bibfield
		{journal} {\bibinfo  {journal} {J. Phys. A: Math. Gen.}\ }\textbf {\bibinfo
			{volume} {29}},\ \bibinfo {pages} {4035} (\bibinfo {year}
		{1996})}\BibitemShut {NoStop}%
	\bibitem [{\citenamefont {Irish}\ \emph {et~al.}(2005)\citenamefont {Irish},
		\citenamefont {Gea-Banacloche}, \citenamefont {Martin},\ and\ \citenamefont
		{Schwab}}]{Irish_2005}%
	\BibitemOpen
	\bibfield  {author} {\bibinfo {author} {\bibfnamefont {E.~K.}\ \bibnamefont
			{Irish}}, \bibinfo {author} {\bibfnamefont {J.}~\bibnamefont
			{Gea-Banacloche}}, \bibinfo {author} {\bibfnamefont {I.}~\bibnamefont
			{Martin}}, \ and\ \bibinfo {author} {\bibfnamefont {K.~C.}\ \bibnamefont
			{Schwab}},\ }\href {\doibase 10.1103/PhysRevB.72.195410} {\bibfield
		{journal} {\bibinfo  {journal} {Phys. Rev. B}\ }\textbf {\bibinfo {volume}
			{72}},\ \bibinfo {pages} {195410} (\bibinfo {year} {2005})}\BibitemShut
	{NoStop}%
	\bibitem [{\citenamefont {Irish}(2007)}]{Irish_2007}%
	\BibitemOpen
	\bibfield  {author} {\bibinfo {author} {\bibfnamefont {E.~K.}\ \bibnamefont
			{Irish}},\ }\href {\doibase 10.1103/PhysRevLett.99.173601} {\bibfield
		{journal} {\bibinfo  {journal} {Phys. Rev. Lett.}\ }\textbf {\bibinfo
			{volume} {99}},\ \bibinfo {pages} {173601} (\bibinfo {year}
		{2007})}\BibitemShut {NoStop}%
	\bibitem [{\citenamefont {Gan}\ and\ \citenamefont {Zheng}(2010)}]{Gan_2010}%
	\BibitemOpen
	\bibfield  {author} {\bibinfo {author} {\bibfnamefont {C.~J.}\ \bibnamefont
			{Gan}}\ and\ \bibinfo {author} {\bibfnamefont {H.}~\bibnamefont {Zheng}},\
	}\href {\doibase 10.1140/epjd/e2010-00182-8} {\bibfield  {journal} {\bibinfo
			{journal} {Eur. Phys. J. D}\ }\textbf {\bibinfo {volume} {59}},\ \bibinfo
		{pages} {473} (\bibinfo {year} {2010})}\BibitemShut {NoStop}%
	\bibitem [{\citenamefont {Boit{\'{e}}}(2020)}]{Boite2020}%
	\BibitemOpen
	\bibfield  {author} {\bibinfo {author} {\bibfnamefont {A.~L.}\ \bibnamefont
			{Boit{\'{e}}}},\ }\href {\doibase 10.1002/qute.201900140} {\bibfield
		{journal} {\bibinfo  {journal} {Adv. Quantum Technol.}\ ,\ \bibinfo {pages}
			{1900140}} (\bibinfo {year} {2020})}\BibitemShut {NoStop}%
	\bibitem [{\citenamefont {Zhang}\ \emph {et~al.}(2011)\citenamefont {Zhang},
		\citenamefont {Chen}, \citenamefont {Yu}, \citenamefont {Liang},
		\citenamefont {Liang},\ and\ \citenamefont {Jia}}]{Zhang_2011}%
	\BibitemOpen
	\bibfield  {author} {\bibinfo {author} {\bibfnamefont {Y.}~\bibnamefont
			{Zhang}}, \bibinfo {author} {\bibfnamefont {G.}~\bibnamefont {Chen}},
		\bibinfo {author} {\bibfnamefont {L.}~\bibnamefont {Yu}}, \bibinfo {author}
		{\bibfnamefont {Q.}~\bibnamefont {Liang}}, \bibinfo {author} {\bibfnamefont
			{J.-Q.}\ \bibnamefont {Liang}}, \ and\ \bibinfo {author} {\bibfnamefont
			{S.}~\bibnamefont {Jia}},\ }\href {\doibase 10.1103/PhysRevA.83.065802}
	{\bibfield  {journal} {\bibinfo  {journal} {Phys. Rev. A}\ }\textbf {\bibinfo
			{volume} {83}},\ \bibinfo {pages} {065802} (\bibinfo {year}
		{2011})}\BibitemShut {NoStop}%
	\bibitem [{\citenamefont {Yu}\ \emph {et~al.}(2012)\citenamefont {Yu},
		\citenamefont {Zhu}, \citenamefont {Liang}, \citenamefont {Chen},\ and\
		\citenamefont {Jia}}]{Yu_2012}%
	\BibitemOpen
	\bibfield  {author} {\bibinfo {author} {\bibfnamefont {L.}~\bibnamefont
			{Yu}}, \bibinfo {author} {\bibfnamefont {S.}~\bibnamefont {Zhu}}, \bibinfo
		{author} {\bibfnamefont {Q.}~\bibnamefont {Liang}}, \bibinfo {author}
		{\bibfnamefont {G.}~\bibnamefont {Chen}}, \ and\ \bibinfo {author}
		{\bibfnamefont {S.}~\bibnamefont {Jia}},\ }\href {\doibase
		10.1103/PhysRevA.86.015803} {\bibfield  {journal} {\bibinfo  {journal} {Phys.
				Rev. A}\ }\textbf {\bibinfo {volume} {86}},\ \bibinfo {pages} {015803}
		(\bibinfo {year} {2012})}\BibitemShut {NoStop}%
	\bibitem [{\citenamefont {Zhang}(2016)}]{Zhang_2016}%
	\BibitemOpen
	\bibfield  {author} {\bibinfo {author} {\bibfnamefont {Y.-Y.}\ \bibnamefont
			{Zhang}},\ }\href {\doibase 10.1103/PhysRevA.94.063824} {\bibfield  {journal}
		{\bibinfo  {journal} {Phys. Rev. A}\ }\textbf {\bibinfo {volume} {94}},\
		\bibinfo {pages} {063824} (\bibinfo {year} {2016})}\BibitemShut {NoStop}%
	\bibitem [{\citenamefont {Zhang}\ and\ \citenamefont
		{Chen}(2017)}]{Zhang_2017}%
	\BibitemOpen
	\bibfield  {author} {\bibinfo {author} {\bibfnamefont {Y.-Y.}\ \bibnamefont
			{Zhang}}\ and\ \bibinfo {author} {\bibfnamefont {X.-Y.}\ \bibnamefont
			{Chen}},\ }\href {\doibase 10.1103/PhysRevA.96.063821} {\bibfield  {journal}
		{\bibinfo  {journal} {Phys. Rev. A}\ }\textbf {\bibinfo {volume} {96}},\
		\bibinfo {pages} {063821} (\bibinfo {year} {2017})}\BibitemShut {NoStop}%
	\bibitem [{\citenamefont {Zhang}\ \emph {et~al.}(2013)\citenamefont {Zhang},
		\citenamefont {Chen},\ and\ \citenamefont {Zhao}}]{Zhang_2013}%
	\BibitemOpen
	\bibfield  {author} {\bibinfo {author} {\bibfnamefont {Y.-Y.}\ \bibnamefont
			{Zhang}}, \bibinfo {author} {\bibfnamefont {Q.-H.}\ \bibnamefont {Chen}}, \
		and\ \bibinfo {author} {\bibfnamefont {Y.}~\bibnamefont {Zhao}},\ }\href
	{\doibase 10.1103/PhysRevA.87.033827} {\bibfield  {journal} {\bibinfo
			{journal} {Phys. Rev. A}\ }\textbf {\bibinfo {volume} {87}},\ \bibinfo
		{pages} {033827} (\bibinfo {year} {2013})}\BibitemShut {NoStop}%
	\bibitem [{\citenamefont {Mao}\ \emph {et~al.}(2018)\citenamefont {Mao},
		\citenamefont {Liu}, \citenamefont {Wu}, \citenamefont {Li}, \citenamefont
		{Ying},\ and\ \citenamefont {Luo}}]{Mao_2018}%
	\BibitemOpen
	\bibfield  {author} {\bibinfo {author} {\bibfnamefont {B.-B.}\ \bibnamefont
			{Mao}}, \bibinfo {author} {\bibfnamefont {M.}~\bibnamefont {Liu}}, \bibinfo
		{author} {\bibfnamefont {W.}~\bibnamefont {Wu}}, \bibinfo {author}
		{\bibfnamefont {L.}~\bibnamefont {Li}}, \bibinfo {author} {\bibfnamefont
			{Z.-J.}\ \bibnamefont {Ying}}, \ and\ \bibinfo {author} {\bibfnamefont
			{H.-G.}\ \bibnamefont {Luo}},\ }\href {\doibase
		10.1088/1674-1056/27/5/054219} {\bibfield  {journal} {\bibinfo  {journal}
			{Chin. Phys. B}\ }\textbf {\bibinfo {volume} {27}},\ \bibinfo {pages}
		{054219} (\bibinfo {year} {2018})}\BibitemShut {NoStop}%
	\bibitem [{\citenamefont {Xie}\ \emph {et~al.}(2020{\natexlab{a}})\citenamefont
		{Xie}, \citenamefont {Mao}, \citenamefont {Li}, \citenamefont {Wang},
		\citenamefont {Sun}, \citenamefont {Wang}, \citenamefont {You},\ and\
		\citenamefont {Liu}}]{Xie2020}%
	\BibitemOpen
	\bibfield  {author} {\bibinfo {author} {\bibfnamefont {W.}~\bibnamefont
			{Xie}}, \bibinfo {author} {\bibfnamefont {B.-B.}\ \bibnamefont {Mao}},
		\bibinfo {author} {\bibfnamefont {G.}~\bibnamefont {Li}}, \bibinfo {author}
		{\bibfnamefont {W.}~\bibnamefont {Wang}}, \bibinfo {author} {\bibfnamefont
			{C.}~\bibnamefont {Sun}}, \bibinfo {author} {\bibfnamefont {Y.}~\bibnamefont
			{Wang}}, \bibinfo {author} {\bibfnamefont {W.-L.}\ \bibnamefont {You}}, \
		and\ \bibinfo {author} {\bibfnamefont {M.}~\bibnamefont {Liu}},\ }\href
	{\doibase 10.1088/1751-8121/ab4b7a} {\bibfield  {journal} {\bibinfo
			{journal} {J. Phys. A: Math. Theor.}\ }\textbf {\bibinfo {volume} {53}},\
		\bibinfo {pages} {095302} (\bibinfo {year} {2020}{\natexlab{a}})}\BibitemShut
	{NoStop}%
	\bibitem [{\citenamefont {Li}\ and\ \citenamefont {Batchelor}(2015)}]{Li_2015}%
	\BibitemOpen
	\bibfield  {author} {\bibinfo {author} {\bibfnamefont {Z.-M.}\ \bibnamefont
			{Li}}\ and\ \bibinfo {author} {\bibfnamefont {M.~T.}\ \bibnamefont
			{Batchelor}},\ }\href {\doibase 10.1088/1751-8113/48/45/454005} {\bibfield
		{journal} {\bibinfo  {journal} {J. Phys. A: Math. Theor.}\ }\textbf {\bibinfo
			{volume} {48}},\ \bibinfo {pages} {454005} (\bibinfo {year}
		{2015})}\BibitemShut {NoStop}%
	\bibitem [{\citenamefont {Li}\ and\ \citenamefont {Batchelor}(2016)}]{Li_2016}%
	\BibitemOpen
	\bibfield  {author} {\bibinfo {author} {\bibfnamefont {Z.-M.}\ \bibnamefont
			{Li}}\ and\ \bibinfo {author} {\bibfnamefont {M.~T.}\ \bibnamefont
			{Batchelor}},\ }\href {\doibase 10.1088/1751-8113/49/36/369401} {\bibfield
		{journal} {\bibinfo  {journal} {J. Phys. A: Math. Theor.}\ }\textbf {\bibinfo
			{volume} {49}},\ \bibinfo {pages} {369401} (\bibinfo {year}
		{2016})}\BibitemShut {NoStop}%
	\bibitem [{\citenamefont {Wakayama}(2017)}]{Wakayama_2017}%
	\BibitemOpen
	\bibfield  {author} {\bibinfo {author} {\bibfnamefont {M.}~\bibnamefont
			{Wakayama}},\ }\href {\doibase 10.1088/1751-8121/aa649b} {\bibfield
		{journal} {\bibinfo  {journal} {J. Phys. A: Math. Theor.}\ }\textbf {\bibinfo
			{volume} {50}},\ \bibinfo {pages} {174001} (\bibinfo {year}
		{2017})}\BibitemShut {NoStop}%
	\bibitem [{\citenamefont {Kimoto}\ \emph {et~al.}(2020)\citenamefont {Kimoto},
		\citenamefont {Reyes-Bustos},\ and\ \citenamefont {Wakayama}}]{Kimoto_2020}%
	\BibitemOpen
	\bibfield  {author} {\bibinfo {author} {\bibfnamefont {K.}~\bibnamefont
			{Kimoto}}, \bibinfo {author} {\bibfnamefont {C.}~\bibnamefont
			{Reyes-Bustos}}, \ and\ \bibinfo {author} {\bibfnamefont {M.}~\bibnamefont
			{Wakayama}},\ }\href {\doibase 10.1093/imrn/rnaa034} {\bibfield  {journal}
		{\bibinfo  {journal} {Int. Math. Res. Not.}\ } (\bibinfo {year} {2020}),\
		10.1093/imrn/rnaa034}\BibitemShut {NoStop}%
	\bibitem [{\citenamefont {Li}\ and\ \citenamefont {Batchelor}()}]{Li2020a}%
	\BibitemOpen
	\bibfield  {author} {\bibinfo {author} {\bibfnamefont {Z.-M.}\ \bibnamefont
			{Li}}\ and\ \bibinfo {author} {\bibfnamefont {M.~T.}\ \bibnamefont
			{Batchelor}},\ }\href@noop {} {\ }\Eprint {http://arxiv.org/abs/2007.06311}
	{arXiv:2007.06311} \BibitemShut {NoStop}%
	\bibitem [{\citenamefont {Li}\ \emph {et~al.}()\citenamefont {Li},
		\citenamefont {Ferri}, \citenamefont {Tilbrook},\ and\ \citenamefont
		{Batchelor}}]{Li2020}%
	\BibitemOpen
	\bibfield  {author} {\bibinfo {author} {\bibfnamefont {Z.-M.}\ \bibnamefont
			{Li}}, \bibinfo {author} {\bibfnamefont {D.}~\bibnamefont {Ferri}}, \bibinfo
		{author} {\bibfnamefont {D.}~\bibnamefont {Tilbrook}}, \ and\ \bibinfo
		{author} {\bibfnamefont {M.~T.}\ \bibnamefont {Batchelor}},\ }\href@noop {}
	{\ }\Eprint {http://arxiv.org/abs/2007.11969} {arXiv:2007.11969} \BibitemShut
	{NoStop}%
	\bibitem [{\citenamefont {Irish}\ and\ \citenamefont
		{Gea-Banacloche}(2014)}]{Irish_2014}%
	\BibitemOpen
	\bibfield  {author} {\bibinfo {author} {\bibfnamefont {E.~K.}\ \bibnamefont
			{Irish}}\ and\ \bibinfo {author} {\bibfnamefont {J.}~\bibnamefont
			{Gea-Banacloche}},\ }\href {\doibase 10.1103/PhysRevB.89.085421} {\bibfield
		{journal} {\bibinfo  {journal} {Phys. Rev. B}\ }\textbf {\bibinfo {volume}
			{89}},\ \bibinfo {pages} {085421} (\bibinfo {year} {2014})}\BibitemShut
	{NoStop}%
	\bibitem [{\citenamefont {Leroux}\ \emph {et~al.}(2017)\citenamefont {Leroux},
		\citenamefont {Govia},\ and\ \citenamefont {Clerk}}]{Leroux_2017}%
	\BibitemOpen
	\bibfield  {author} {\bibinfo {author} {\bibfnamefont {C.}~\bibnamefont
			{Leroux}}, \bibinfo {author} {\bibfnamefont {L.~C.~G.}\ \bibnamefont
			{Govia}}, \ and\ \bibinfo {author} {\bibfnamefont {A.~A.}\ \bibnamefont
			{Clerk}},\ }\href {\doibase 10.1103/PhysRevA.96.043834} {\bibfield  {journal}
		{\bibinfo  {journal} {Phys. Rev. A}\ }\textbf {\bibinfo {volume} {96}},\
		\bibinfo {pages} {043834} (\bibinfo {year} {2017})}\BibitemShut {NoStop}%
	\bibitem [{\citenamefont {Ying}\ \emph {et~al.}(2015)\citenamefont {Ying},
		\citenamefont {Liu}, \citenamefont {Luo}, \citenamefont {Lin},\ and\
		\citenamefont {You}}]{Ying_2015}%
	\BibitemOpen
	\bibfield  {author} {\bibinfo {author} {\bibfnamefont {Z.-J.}\ \bibnamefont
			{Ying}}, \bibinfo {author} {\bibfnamefont {M.}~\bibnamefont {Liu}}, \bibinfo
		{author} {\bibfnamefont {H.-G.}\ \bibnamefont {Luo}}, \bibinfo {author}
		{\bibfnamefont {H.-Q.}\ \bibnamefont {Lin}}, \ and\ \bibinfo {author}
		{\bibfnamefont {J.~Q.}\ \bibnamefont {You}},\ }\href {\doibase
		10.1103/PhysRevA.92.053823} {\bibfield  {journal} {\bibinfo  {journal} {Phys.
				Rev. A}\ }\textbf {\bibinfo {volume} {92}},\ \bibinfo {pages} {053823}
		(\bibinfo {year} {2015})}\BibitemShut {NoStop}%
	\bibitem [{\citenamefont {Ashhab}(2020)}]{Ashhab_2020}%
	\BibitemOpen
	\bibfield  {author} {\bibinfo {author} {\bibfnamefont {S.}~\bibnamefont
			{Ashhab}},\ }\href {\doibase 10.1103/PhysRevA.101.023808} {\bibfield
		{journal} {\bibinfo  {journal} {Phys. Rev. A}\ }\textbf {\bibinfo {volume}
			{101}},\ \bibinfo {pages} {023808} (\bibinfo {year} {2020})}\BibitemShut
	{NoStop}%
	\bibitem [{\citenamefont {Hwang}\ \emph {et~al.}(2015)\citenamefont {Hwang},
		\citenamefont {Puebla},\ and\ \citenamefont {Plenio}}]{Hwang_2015}%
	\BibitemOpen
	\bibfield  {author} {\bibinfo {author} {\bibfnamefont {M.-J.}\ \bibnamefont
			{Hwang}}, \bibinfo {author} {\bibfnamefont {R.}~\bibnamefont {Puebla}}, \
		and\ \bibinfo {author} {\bibfnamefont {M.~B.}\ \bibnamefont {Plenio}},\
	}\href {\doibase 10.1103/PhysRevLett.115.180404} {\bibfield  {journal}
		{\bibinfo  {journal} {Phys. Rev. Lett.}\ }\textbf {\bibinfo {volume} {115}},\
		\bibinfo {pages} {180404} (\bibinfo {year} {2015})}\BibitemShut {NoStop}%
	\bibitem [{\citenamefont {Chen}\ \emph {et~al.}()\citenamefont {Chen},
		\citenamefont {Xie},\ and\ \citenamefont {Chen}}]{Chen2020}%
	\BibitemOpen
	\bibfield  {author} {\bibinfo {author} {\bibfnamefont {X.-Y.}\ \bibnamefont
			{Chen}}, \bibinfo {author} {\bibfnamefont {Y.-F.}\ \bibnamefont {Xie}}, \
		and\ \bibinfo {author} {\bibfnamefont {Q.-H.}\ \bibnamefont {Chen}},\
	}\href@noop {} {\ }\Eprint {http://arxiv.org/abs/2001.04356}
	{arXiv:2001.04356} \BibitemShut {NoStop}%
	\bibitem [{\citenamefont {Hwang}\ and\ \citenamefont
		{Choi}(2010)}]{Hwang_2010}%
	\BibitemOpen
	\bibfield  {author} {\bibinfo {author} {\bibfnamefont {M.-J.}\ \bibnamefont
			{Hwang}}\ and\ \bibinfo {author} {\bibfnamefont {M.-S.}\ \bibnamefont
			{Choi}},\ }\href {\doibase 10.1103/PhysRevA.82.025802} {\bibfield  {journal}
		{\bibinfo  {journal} {Phys. Rev. A}\ }\textbf {\bibinfo {volume} {82}},\
		\bibinfo {pages} {025802} (\bibinfo {year} {2010})}\BibitemShut {NoStop}%
	\bibitem [{\citenamefont {Fedorov}\ \emph {et~al.}(2010)\citenamefont
		{Fedorov}, \citenamefont {Feofanov}, \citenamefont {Macha}, \citenamefont
		{Forn-D\'{\i}az}, \citenamefont {Harmans},\ and\ \citenamefont
		{Mooij}}]{Fedorov_2010}%
	\BibitemOpen
	\bibfield  {author} {\bibinfo {author} {\bibfnamefont {A.}~\bibnamefont
			{Fedorov}}, \bibinfo {author} {\bibfnamefont {A.~K.}\ \bibnamefont
			{Feofanov}}, \bibinfo {author} {\bibfnamefont {P.}~\bibnamefont {Macha}},
		\bibinfo {author} {\bibfnamefont {P.}~\bibnamefont {Forn-D\'{\i}az}},
		\bibinfo {author} {\bibfnamefont {C.~J. P.~M.}\ \bibnamefont {Harmans}}, \
		and\ \bibinfo {author} {\bibfnamefont {J.~E.}\ \bibnamefont {Mooij}},\ }\href
	{\doibase 10.1103/PhysRevLett.105.060503} {\bibfield  {journal} {\bibinfo
			{journal} {Phys. Rev. Lett.}\ }\textbf {\bibinfo {volume} {105}},\ \bibinfo
		{pages} {060503} (\bibinfo {year} {2010})}\BibitemShut {NoStop}%
	\bibitem [{Note1()}]{Note1}%
	\BibitemOpen
	\bibinfo {note} {For example, in \protect \textit {Mathematica 12} we simply
		use the function \protect \textbf {NMinimize} to minimize the ground state
		energy defined in Eq.~(\ref {Egs})}\BibitemShut {NoStop}%
	\bibitem [{\citenamefont {Chen}\ \emph {et~al.}(2020)\citenamefont {Chen},
		\citenamefont {Zhang}, \citenamefont {Fu},\ and\ \citenamefont
		{Zheng}}]{Chen_2020}%
	\BibitemOpen
	\bibfield  {author} {\bibinfo {author} {\bibfnamefont {X.-Y.}\ \bibnamefont
			{Chen}}, \bibinfo {author} {\bibfnamefont {Y.-Y.}\ \bibnamefont {Zhang}},
		\bibinfo {author} {\bibfnamefont {L.}~\bibnamefont {Fu}}, \ and\ \bibinfo
		{author} {\bibfnamefont {H.}~\bibnamefont {Zheng}},\ }\href {\doibase
		10.1103/PhysRevA.101.033827} {\bibfield  {journal} {\bibinfo  {journal}
			{Phys. Rev. A}\ }\textbf {\bibinfo {volume} {101}},\ \bibinfo {pages}
		{033827} (\bibinfo {year} {2020})}\BibitemShut {NoStop}%
	\bibitem [{\citenamefont {Liu}\ \emph {et~al.}(2017)\citenamefont {Liu},
		\citenamefont {Ying}, \citenamefont {An}, \citenamefont {Luo},\ and\
		\citenamefont {Lin}}]{Liu2017a}%
	\BibitemOpen
	\bibfield  {author} {\bibinfo {author} {\bibfnamefont {M.}~\bibnamefont
			{Liu}}, \bibinfo {author} {\bibfnamefont {Z.-J.}\ \bibnamefont {Ying}},
		\bibinfo {author} {\bibfnamefont {J.-H.}\ \bibnamefont {An}}, \bibinfo
		{author} {\bibfnamefont {H.-G.}\ \bibnamefont {Luo}}, \ and\ \bibinfo
		{author} {\bibfnamefont {H.-Q.}\ \bibnamefont {Lin}},\ }\href {\doibase
		10.1088/1751-8121/aa56f6} {\bibfield  {journal} {\bibinfo  {journal} {J.
				Phys. A: Math. Theor.}\ }\textbf {\bibinfo {volume} {50}},\ \bibinfo {pages}
		{084003} (\bibinfo {year} {2017})}\BibitemShut {NoStop}%
	\bibitem [{\citenamefont {Cong}\ \emph {et~al.}(2017)\citenamefont {Cong},
		\citenamefont {Sun}, \citenamefont {Liu}, \citenamefont {Ying},\ and\
		\citenamefont {Luo}}]{Cong_2017}%
	\BibitemOpen
	\bibfield  {author} {\bibinfo {author} {\bibfnamefont {L.}~\bibnamefont
			{Cong}}, \bibinfo {author} {\bibfnamefont {X.-M.}\ \bibnamefont {Sun}},
		\bibinfo {author} {\bibfnamefont {M.}~\bibnamefont {Liu}}, \bibinfo {author}
		{\bibfnamefont {Z.-J.}\ \bibnamefont {Ying}}, \ and\ \bibinfo {author}
		{\bibfnamefont {H.-G.}\ \bibnamefont {Luo}},\ }\href {\doibase
		10.1103/PhysRevA.95.063803} {\bibfield  {journal} {\bibinfo  {journal} {Phys.
				Rev. A}\ }\textbf {\bibinfo {volume} {95}},\ \bibinfo {pages} {063803}
		(\bibinfo {year} {2017})}\BibitemShut {NoStop}%
	\bibitem [{\citenamefont {Cong}\ \emph {et~al.}(2019)\citenamefont {Cong},
		\citenamefont {Sun}, \citenamefont {Liu}, \citenamefont {Ying},\ and\
		\citenamefont {Luo}}]{Cong_2019}%
	\BibitemOpen
	\bibfield  {author} {\bibinfo {author} {\bibfnamefont {L.}~\bibnamefont
			{Cong}}, \bibinfo {author} {\bibfnamefont {X.-M.}\ \bibnamefont {Sun}},
		\bibinfo {author} {\bibfnamefont {M.}~\bibnamefont {Liu}}, \bibinfo {author}
		{\bibfnamefont {Z.-J.}\ \bibnamefont {Ying}}, \ and\ \bibinfo {author}
		{\bibfnamefont {H.-G.}\ \bibnamefont {Luo}},\ }\href {\doibase
		10.1103/PhysRevA.99.013815} {\bibfield  {journal} {\bibinfo  {journal} {Phys.
				Rev. A}\ }\textbf {\bibinfo {volume} {99}},\ \bibinfo {pages} {013815}
		(\bibinfo {year} {2019})}\BibitemShut {NoStop}%
	\bibitem [{\citenamefont {Sun}\ \emph {et~al.}(2020)\citenamefont {Sun},
		\citenamefont {Cong}, \citenamefont {Eckle}, \citenamefont {Ying},\ and\
		\citenamefont {Luo}}]{Sun2019}%
	\BibitemOpen
	\bibfield  {author} {\bibinfo {author} {\bibfnamefont {X.-M.}\ \bibnamefont
			{Sun}}, \bibinfo {author} {\bibfnamefont {L.}~\bibnamefont {Cong}}, \bibinfo
		{author} {\bibfnamefont {H.-P.}\ \bibnamefont {Eckle}}, \bibinfo {author}
		{\bibfnamefont {Z.-J.}\ \bibnamefont {Ying}}, \ and\ \bibinfo {author}
		{\bibfnamefont {H.-G.}\ \bibnamefont {Luo}},\ }\href {\doibase
		10.1103/PhysRevA.101.063832} {\bibfield  {journal} {\bibinfo  {journal}
			{Phys. Rev. A}\ }\textbf {\bibinfo {volume} {101}},\ \bibinfo {pages}
		{063832} (\bibinfo {year} {2020})}\BibitemShut {NoStop}%
	\bibitem [{\citenamefont {Tomka}\ \emph {et~al.}(2014)\citenamefont {Tomka},
		\citenamefont {El~Araby}, \citenamefont {Pletyukhov},\ and\ \citenamefont
		{Gritsev}}]{Tomka_2014}%
	\BibitemOpen
	\bibfield  {author} {\bibinfo {author} {\bibfnamefont {M.}~\bibnamefont
			{Tomka}}, \bibinfo {author} {\bibfnamefont {O.}~\bibnamefont {El~Araby}},
		\bibinfo {author} {\bibfnamefont {M.}~\bibnamefont {Pletyukhov}}, \ and\
		\bibinfo {author} {\bibfnamefont {V.}~\bibnamefont {Gritsev}},\ }\href
	{\doibase 10.1103/PhysRevA.90.063839} {\bibfield  {journal} {\bibinfo
			{journal} {Phys. Rev. A}\ }\textbf {\bibinfo {volume} {90}},\ \bibinfo
		{pages} {063839} (\bibinfo {year} {2014})}\BibitemShut {NoStop}%
	\bibitem [{\citenamefont {Xie}\ \emph {et~al.}(2014)\citenamefont {Xie},
		\citenamefont {Cui}, \citenamefont {Cao}, \citenamefont {Amico},\ and\
		\citenamefont {Fan}}]{Xie_2014}%
	\BibitemOpen
	\bibfield  {author} {\bibinfo {author} {\bibfnamefont {Q.-T.}\ \bibnamefont
			{Xie}}, \bibinfo {author} {\bibfnamefont {S.}~\bibnamefont {Cui}}, \bibinfo
		{author} {\bibfnamefont {J.-P.}\ \bibnamefont {Cao}}, \bibinfo {author}
		{\bibfnamefont {L.}~\bibnamefont {Amico}}, \ and\ \bibinfo {author}
		{\bibfnamefont {H.}~\bibnamefont {Fan}},\ }\href {\doibase
		10.1103/PhysRevX.4.021046} {\bibfield  {journal} {\bibinfo  {journal} {Phys.
				Rev. X}\ }\textbf {\bibinfo {volume} {4}},\ \bibinfo {pages} {021046}
		(\bibinfo {year} {2014})}\BibitemShut {NoStop}%
	\bibitem [{\citenamefont {Eckle}\ and\ \citenamefont
		{Johannesson}(2017)}]{Eckle_2017}%
	\BibitemOpen
	\bibfield  {author} {\bibinfo {author} {\bibfnamefont {H.-P.}\ \bibnamefont
			{Eckle}}\ and\ \bibinfo {author} {\bibfnamefont {H.}~\bibnamefont
			{Johannesson}},\ }\href {\doibase 10.1088/1751-8121/aa785a} {\bibfield
		{journal} {\bibinfo  {journal} {J. Phys. A: Math. Theor.}\ }\textbf {\bibinfo
			{volume} {50}},\ \bibinfo {pages} {294004} (\bibinfo {year}
		{2017})}\BibitemShut {NoStop}%
	\bibitem [{\citenamefont {Xie}\ \emph {et~al.}(2019)\citenamefont {Xie},
		\citenamefont {Duan},\ and\ \citenamefont {Chen}}]{Xie_2019}%
	\BibitemOpen
	\bibfield  {author} {\bibinfo {author} {\bibfnamefont {Y.-F.}\ \bibnamefont
			{Xie}}, \bibinfo {author} {\bibfnamefont {L.}~\bibnamefont {Duan}}, \ and\
		\bibinfo {author} {\bibfnamefont {Q.-H.}\ \bibnamefont {Chen}},\ }\href
	{\doibase 10.1088/1751-8121/ab1cf6} {\bibfield  {journal} {\bibinfo
			{journal} {J. Phys. A: Math. Theor.}\ }\textbf {\bibinfo {volume} {52}},\
		\bibinfo {pages} {245304} (\bibinfo {year} {2019})}\BibitemShut {NoStop}%
	\bibitem [{\citenamefont {Cong}\ \emph {et~al.}(2020)\citenamefont {Cong},
		\citenamefont {Felicetti}, \citenamefont {Casanova}, \citenamefont {Lamata},
		\citenamefont {Solano},\ and\ \citenamefont {Arrazola}}]{Cong_2020}%
	\BibitemOpen
	\bibfield  {author} {\bibinfo {author} {\bibfnamefont {L.}~\bibnamefont
			{Cong}}, \bibinfo {author} {\bibfnamefont {S.}~\bibnamefont {Felicetti}},
		\bibinfo {author} {\bibfnamefont {J.}~\bibnamefont {Casanova}}, \bibinfo
		{author} {\bibfnamefont {L.}~\bibnamefont {Lamata}}, \bibinfo {author}
		{\bibfnamefont {E.}~\bibnamefont {Solano}}, \ and\ \bibinfo {author}
		{\bibfnamefont {I.}~\bibnamefont {Arrazola}},\ }\href {\doibase
		10.1103/PhysRevA.101.032350} {\bibfield  {journal} {\bibinfo  {journal}
			{Phys. Rev. A}\ }\textbf {\bibinfo {volume} {101}},\ \bibinfo {pages}
		{032350} (\bibinfo {year} {2020})}\BibitemShut {NoStop}%
	\bibitem [{\citenamefont {Xie}\ \emph {et~al.}(2020{\natexlab{b}})\citenamefont
		{Xie}, \citenamefont {Chen}, \citenamefont {Dong},\ and\ \citenamefont
		{Chen}}]{Xie_2020}%
	\BibitemOpen
	\bibfield  {author} {\bibinfo {author} {\bibfnamefont {Y.-F.}\ \bibnamefont
			{Xie}}, \bibinfo {author} {\bibfnamefont {X.-Y.}\ \bibnamefont {Chen}},
		\bibinfo {author} {\bibfnamefont {X.-F.}\ \bibnamefont {Dong}}, \ and\
		\bibinfo {author} {\bibfnamefont {Q.-H.}\ \bibnamefont {Chen}},\ }\href
	{\doibase 10.1103/PhysRevA.101.053803} {\bibfield  {journal} {\bibinfo
			{journal} {Phys. Rev. A}\ }\textbf {\bibinfo {volume} {101}},\ \bibinfo
		{pages} {053803} (\bibinfo {year} {2020}{\natexlab{b}})}\BibitemShut
	{NoStop}%
		\bibitem [{\citenamefont {Batchelor}\ \emph {et~al.}(2016)\citenamefont
		{Batchelor}, \citenamefont {Li},\ and\ \citenamefont {Zhou}}]{Batchelor2015}%
	\BibitemOpen
	\bibfield  {author} {\bibinfo {author} {\bibfnamefont {M.~T.}\ \bibnamefont
			{Batchelor}}, \bibinfo {author} {\bibfnamefont {Z.-M.}\ \bibnamefont {Li}}, \
		and\ \bibinfo {author} {\bibfnamefont {H.-Q.}\ \bibnamefont {Zhou}},\ }\href
	{\doibase 10.1088/1751-8113/49/1/01lt01} {\bibfield  {journal} {\bibinfo
			{journal} {J. Phys. A: Math. Theor.}\ }\textbf {\bibinfo {volume} {49}},\
		\bibinfo {pages} {01LT01} (\bibinfo {year} {2016})}\BibitemShut {NoStop}%

\end{thebibliography}

%

\end{document}